\documentclass[10pt,journal,compsoc]{IEEEtran}

\usepackage{hyperref}       
\usepackage{flushend}       
\usepackage{todonotes}      
\usepackage{setspace}       
\usepackage{url}            
\usepackage{booktabs}       
\usepackage{amsfonts}       
\usepackage{nicefrac}       
\usepackage{microtype}      
\usepackage{graphicx}
\usepackage{multirow}
\usepackage{graphicx}
\usepackage{caption}
\usepackage{amssymb}
\usepackage{pifont}
\usepackage{color}
\usepackage{bm}
\usepackage{float}
\usepackage{colortbl}

\ifCLASSOPTIONcompsoc
  \usepackage[nocompress]{cite}
\else
  \usepackage{cite}
\fi
\ifCLASSINFOpdf
\else
\fi

\begin{document}
\title{Pixel-Inconsistency Modeling for Image Manipulation Localization}
\author{Chenqi~Kong,~\IEEEmembership{Member,~IEEE,}
        Anwei~Luo,       Shiqi~Wang,~\IEEEmembership{Senior~Member,~IEEE,}
        Haoliang~Li,~\IEEEmembership{Member,~IEEE,}
        Anderson~Rocha,~\IEEEmembership{Fellow,~IEEE,} and~Alex~C.~Kot,~\IEEEmembership{Life~Fellow,~IEEE}

\IEEEcompsocitemizethanks{\IEEEcompsocthanksitem {C. Kong and A. Kot are with the Rapid-Rich Object Search (ROSE) Lab, School of Electrical and Electronic Engineering, Nanyang Technology University, Singapore, 639798. \protect\\ 
E-mail: chenqi.kong@ntu.edu.sg, eackot@ntu.edu.sg.}
\IEEEcompsocthanksitem {A. Luo is with the School of Computer Science and Engineering, Sun Yat-sen University, Guangzhou, China. He is also with the Rapid-Rich Object Search (ROSE) Lab, School of Electrical and Electronic Engineering, Nanyang Technology University, Singapore, 639798. \protect\\ 
E-mail: luoanw@mail2.sysu.edu.cn.}
\IEEEcompsocthanksitem {S. Wang is with the Department of Computer Science, City University of Hong Kong, Hong Kong. \protect\\ 
E-mail: shiqwang@cityu.edu.hk.}
\IEEEcompsocthanksitem {H. Li is with the Department of Electrical Engineering, City University of Hong Kong, Hong Kong. \protect\\ 
E-mail: haoliang.li@cityu.edu.hk.}
\IEEEcompsocthanksitem {A. Rocha is with the Artificial Intelligence Laboratory (Recod.ai), Institute of Computing, University of Campinas, Campinas 13084-851, Brazil \protect\\ E-mail: arrocha@unicamp.br}
\IEEEcompsocthanksitem {Corresponding author: Haoliang Li.}
}}

\markboth{Journal of \LaTeX\ Class Files,~Vol.~14, No.~8, August~2015}%
{Shell \MakeLowercase{\textit{et al.}}: Bare Demo of IEEEtran.cls for Computer Society Journals}

\IEEEtitleabstractindextext{%
\begin{abstract}
Digital image forensics plays a crucial role in image authentication and manipulation localization. Despite the  progress powered by deep neural networks, existing forgery localization methodologies exhibit limitations when deployed to unseen datasets and perturbed images (i.e., lack of generalization and robustness to real-world applications). To circumvent these problems and aid image integrity, this paper presents a  generalized and robust manipulation localization model through the analysis of pixel inconsistency artifacts. The rationale is grounded on the observation that most image signal processors (ISP) involve the demosaicing process, which introduces pixel correlations in pristine images. Moreover, manipulating operations, including splicing, copy-move, and inpainting, directly affect such pixel regularity. We, therefore, first split the input image into several blocks and design masked self-attention mechanisms to model the global pixel dependency in input images. Simultaneously, we optimize another local pixel dependency stream to mine local manipulation clues within input forgery images. In addition, we design novel Learning-to-Weight Modules (LWM) to combine features from the two streams, thereby enhancing the final forgery localization performance. To improve the training process, we propose a novel Pixel-Inconsistency Data Augmentation (PIDA) strategy, driving the model to focus on capturing inherent pixel-level artifacts instead of mining semantic forgery traces. This work establishes a comprehensive benchmark integrating 16 representative detection models across 12 datasets. Extensive experiments show that our method successfully extracts inherent pixel-inconsistency forgery fingerprints and achieve state-of-the-art generalization and robustness performances in image manipulation localization.
\end{abstract}

}
\maketitle

\IEEEdisplaynontitleabstractindextext
\IEEEpeerreviewmaketitle

\IEEEraisesectionheading{\section{Introduction}\label{sec:introduction}}

\begin{figure}[ht]
\centering
\includegraphics[scale=0.32]{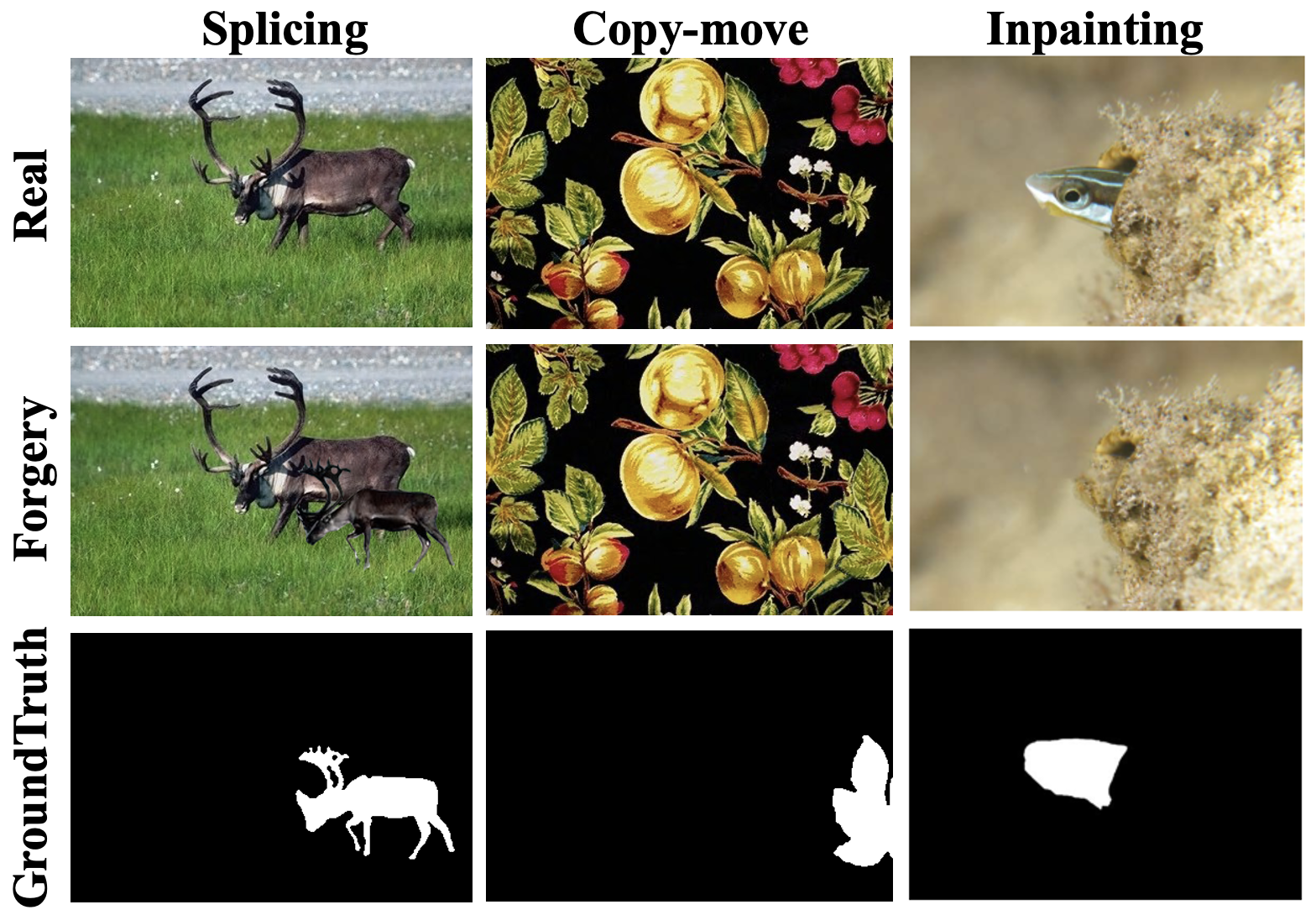}
\caption{Illustration of manipulation types: splicing, copy-move, and inpainting. The top, middle, and bottom rows show the real, forgery, and ground-truth images.}
\label{showcase_dist}
\end{figure}

\IEEEPARstart{I}{mage} manipulation has been carried out since photography was born \cite{image_mani}. In recent decades, there has been significant advances in image manipulation techniques, including splicing, copy-move, and inpainting, which are three pervasive but notorious attack types \cite{verdoliva2020media}, as shown in Fig.~\ref{showcase_dist}. These techniques can produce forgery content with a very high level of realism, blurring the boundaries between authentic and forgery images. Manipulation traces are very subtle and can hardly be perceived by the naked eye. With the widespread use of digital images on the internet, it has become much easier for malicious attackers to launch manipulation attacks using off-the-shelf yet powerful image editing tools, such as Photoshop, After Effects Pro, GIMP, and more recently, Firefly. The produced sophisticated content can be used to commit fraud, generate fake news, and blackmail people. Image manipulation certainly undermines the trust in media content. Moreover, the proliferation of fakes has raised pressing security concerns for the public. Therefore, designing effective image forgery localization models to address these issues is paramount.

Early attempts at image manipulation localization mainly focused on extracting features based on prior knowledge, such as lens distortions \cite{mayer2018accurate, yerushalmy2011digital, johnson2006exposing, yerushalmy2011digital, gloe2010efficient, fu2012forgery}, Color Filter Array (CFA) artifacts \cite{cao2009accurate, ferrara2012image, popescu2005exposing, gallagher2008image, ho2010inter}, noise patterns \cite{lyu2014exposing, kobayashi2010detecting, popescu2004statistical, mahdian2009using, cozzolino2015splicebuster, fan2013estimating}, compression artifacts \cite{fan2003identification, chen2011detecting, iakovidou2018content, barni2010identification, bianchi2012image, fu2007generalized, pasquini2017statistical}. However, these traditional methods demonstrate limited accuracy and generalizability. In turn, learning-based detectors have been proposed thanks to recent advancements in deep learning and artificial intelligence. These methods exhibit promising performance in image forgery localization under the intra-domain setting. Nonetheless, data-driven methods are typically prone to overfitting the training data, resulting in limited robustness and generalization performance. Namely, they are fragile to image perturbations and vulnerable to unseen image manipulation datasets. 

\begin{figure}[ht]
\centering
\includegraphics[scale=0.24]{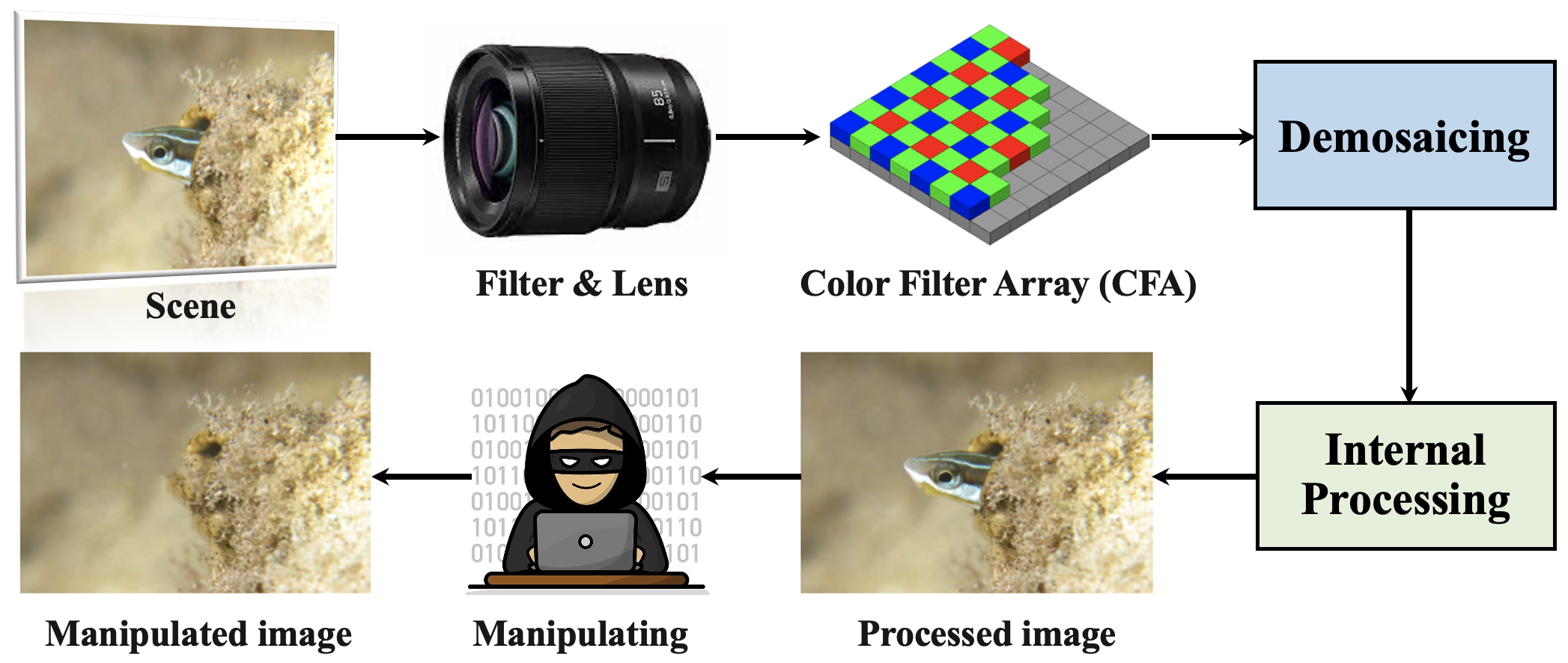}
\caption{Typical forgery image construction pipeline.}
\label{demosaicing}
\end{figure}

Extracting inherent forgery fingerprints for generalized and robust image forgery localization remains a challenging problem. This paper recasts the typical image manipulation pipeline and proposes a new forgery localization framework that captures the pixel inconsistencies in manipulated images. Fig.~\ref{demosaicing} shows the typical forgery image construction chain. The filter and lens eliminate undesired light and focus light onto the sensor. Subsequently, the Color Filter Array (CFA) is applied to extract single-color components. A series of software operations is carried out during the in-camera processing. Demosaicing, also known as color interpolation, is performed to reconstruct full-color pixels from surrounding single-color pixels. Some internal processing steps, such as color correction, noise reduction, and compression, are subsequently conducted to generate the final processed RGB image. In turn, malicious attackers can utilize image editing tools to manipulate pristine images during the out-camera processing. These manipulations can disrupt such pixel correlation (i.e., perturb the periodic patterns) introduced by the demosaicing operation, leaving distinctive pixel inconsistency artifacts for forensics analyses \cite{cao2009accurate, popescu2005exposing, verdoliva2020media}. 

\begin{figure}[ht]
\centering
\includegraphics[scale=0.30]{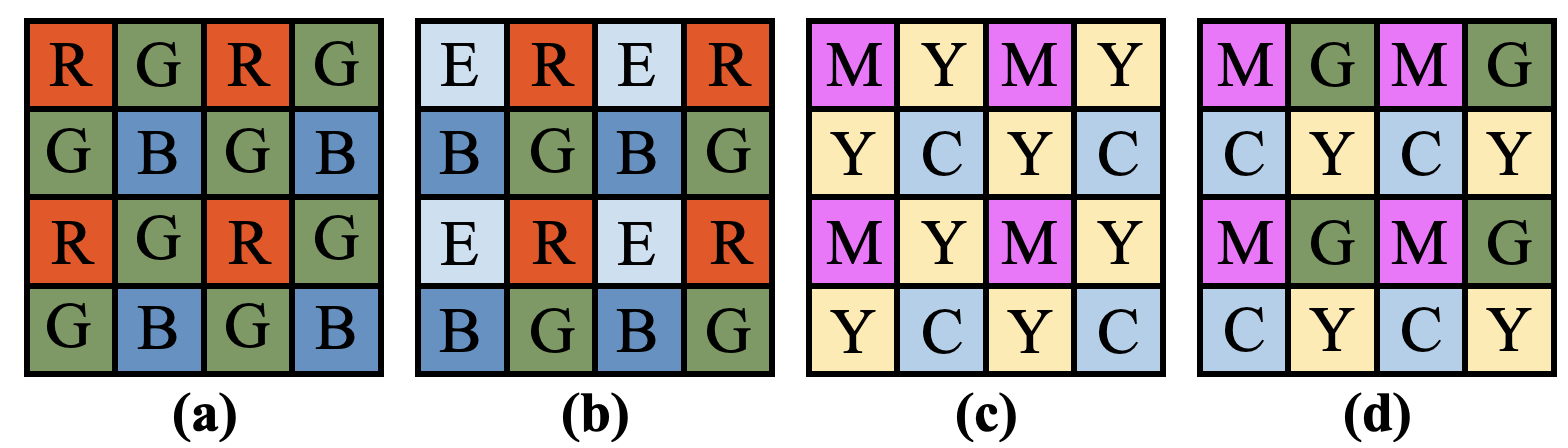}
\caption{Typical Color Filter Array (CFA) types. (a). Bayer CFA; (b). RGBE; (c). CMY; (d). CMYG.}
\label{CFA}
\end{figure}

Fig.~\ref{CFA} showcases four typical CFA types: (a). Bayer CFA, (b). RGBE, (c). CMY, and (d). CMYG. Color filtering allows the capture of a specific color at each pixel. Consequently, in the resulting RAW image, only one color is present at each pixel, and the demosaicing process reconstructs the missing color samples. Some existing forensics analysis techniques for forgery fingerprint extraction focus on mathematically modeling different image regularities. For instance, Popescu $et~al.$ \cite{popescu2005exposing} quantifies the specific correlations introduced by CFA interpolation and describes how these correlations can be automatically detected. Ferrara $et~al.$ \cite{ferrara2012image} proposes a novel feature that measures the presence or absence of these image regularities at the smallest 2$\times$2 block level, thus predicting a forgery probability map. In \cite{cao2009accurate2} and \cite{li2016color}, the intra-block fingerprint is modeled using a linear regression approach. Despite the effectiveness of these pixel correlation modeling approaches in forensic analysis, most require knowledge of the CFA type as prior information. Furthermore, these methods cannot sufficiently capture more complex regularities introduced by smart image signal processors (ISPs) in modern AI cameras \cite{AIcamera}. 




Different from the prior arts, we propose a learning-based method to capture inherent pixel inconsistencies within forged images based on this insight. We design a two-stream pixel-dependency modeling framework for image manipulation localization to achieve this. Drawing inspiration from recent success of autoregressive models (e.g., PixelCNN \cite{van2016pixel, van2016conditional}) in various computer vision tasks, we design a masked self-attention mechanism to model the global pixel dependency within input images. Furthermore, we design a Difference Convolution (DC) stream to capture local pixel inconsistency artifacts within local image regions. In addition, we introduce a novel Learning-to-Weight Modules (LWM) to combine global and local pixel-inconsistency features from these two streams.

We design three decoders to predict the potential manipulated regions, forgery boundaries, and reconstructed images. We finally introduce the Pixel-Inconsistency Data Augmentation (PIDA) strategy to explore the pixel-level forgery traces. PIDA is an effective approach that relies upon only real images for data augmentation. It guides the model to focus on capturing pixel-inconsistency artifacts rather than semantic forgery traces. The designed framework is trained end-to-end, jointly supervised by the binary mask and boundary labels.

\begin{figure*}[ht]
\centering
\includegraphics[scale=0.325]{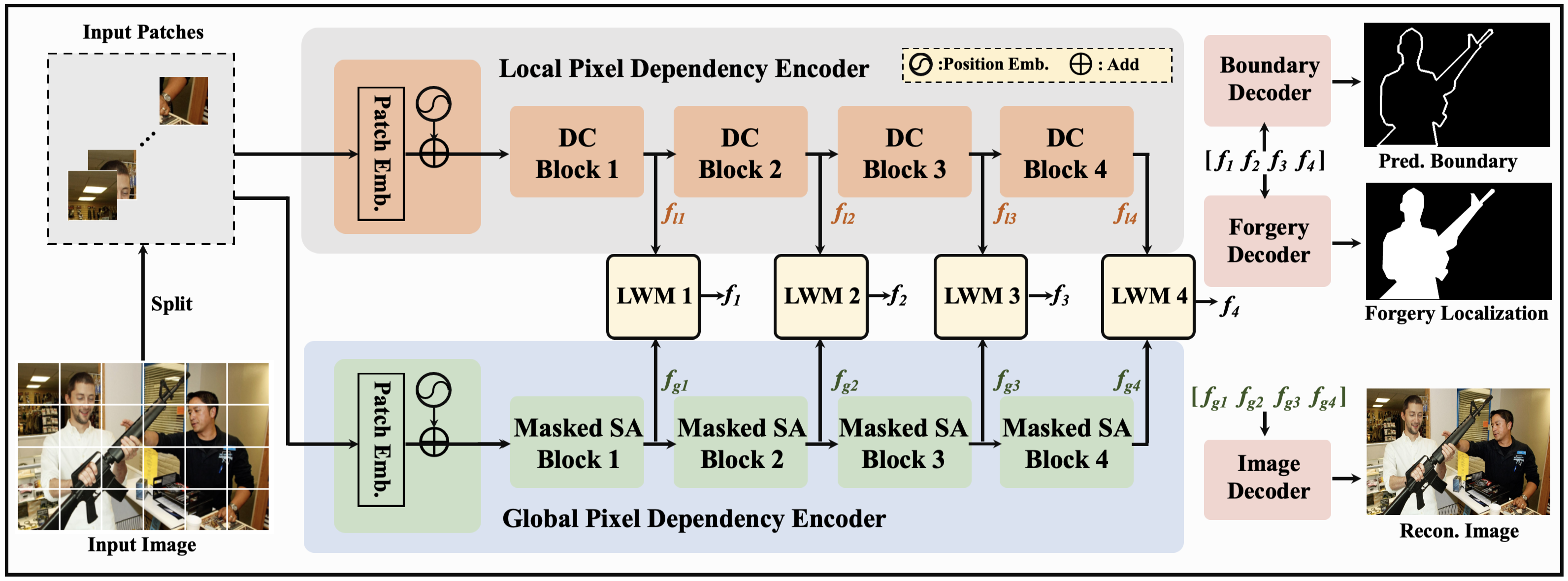}
\caption{Proposed image manipulation localization framework. The input image is split into several patches, which are simultaneously fed forward to the 
Local Pixel Dependency Encoder and Global Pixel Dependency Encoder. The upper stream comprises four Difference Convolution (DC) blocks to capture local pixel inconsistencies in forged images. Meanwhile, the Global Pixel Dependency Encoder, which incorporates four masked self-attention (Masked SA) blocks, focuses on modeling long-range statistics within the input images. Four Learning-to-Weight Modules (LWM) have been devised to combine global and local features extracted by the two encoders. The Forgery Decoder and Boundary Decoder take the aggregated features as inputs and predict the final forgery and boundary maps.  
}
\label{framework}
\end{figure*}

The key contributions of our work are:
\begin{itemize}
    \item We establish a comprehensive benchmark assessing the generalization capabilities of 16 representative image forgery localization methods across 12 datasets. We further extend this benchmark to evaluate the robustness performance across six unseen image perturbation types, each with nine severity levels. \textcolor{black}{Additionally, we evaluate our designed model on sophisticated and advanced manipulations generated by modern Artificial Intelligence Generated Content (AIGC) techniques.}
    \item We design a two-stream image manipulation localization framework comprising a local pixel dependency encoder, a global pixel dependency encoder, four feature fusion modules, and three decoders. The proposed model can effectively extract the pixel-inconsistency forgery fingerprints, leading to more generalized and robust manipulation localization performance.  
    \item We introduce a Pixel-Inconsistency Data Augmentation strategy that exclusively utilizes real images to create the generated data. The proposed data augmentation drives the model to focus on capturing the inherent pixel-level artifacts rather than the semantic forgery clues, contributing to a forgery localization performance boost. 
    \item Extensive quantitative and qualitative experimental results demonstrate that our proposed method consistently outperforms state-of-the-art in generalization and robustness evaluations. \textcolor{black}{Comprehensive ablation experiments further illustrate the effectiveness of the designed components. }
\end{itemize}

Sec. 2 overviews prior work in image forgery localization and pixel dependency modeling. Sec. 3  elaborates on the designed framework. Sec. 4 presents comprehensive evaluation results under diverse experimental settings. Finally, Sec. 5 concludes this paper and discusses current limitations and possible future research directions.


\section{Related Work}
In this section, we broadly review existing works on image forgery detection and localization, including both handcrafted and learning-based methodologies. Additionally, we review the studies related to pixel dependency modeling and their applications. 

\subsection{Manipulation detection and localization methods using low-level traces}  Image manipulation detection is no new problem. Early methods focus on detecting low-level artifacts derived from in-camera processing traces. For example, lens distortions \cite{mayer2018accurate, yerushalmy2011digital, johnson2006exposing, yerushalmy2011digital, gloe2010efficient, fu2012forgery}, introduced by the imperfection of complex optical systems, can be regarded as unique fingerprints for forensics purposes. Chromatic aberration is a typical lens distortion cue widely studied for forgery detection \cite{johnson2006exposing, yerushalmy2011digital, mayer2018accurate}. Besides, many methods \cite{ferrara2012image, popescu2005exposing, cao2009accurate, cao2009accurate2} propose to capture color filter array (CFA) artifacts to detect manipulations. These techniques demonstrated that manipulation operations can disrupt periodic patterns introduced by the demosaicing process. Additionally, since photo-response nonuniformity (PRNU) is specific to each camera model, some methods \cite{lyu2014exposing, mahdian2009using, kobayashi2010detecting} extract noise patterns from query images for detecting digital tampering traces. 
Furthermore, extensive research has been dedicated to studying JPEG compression artifacts that persist in the discrete cosine transform (DCT) domain \cite{chen2011detecting, farid2009exposing, fan2003identification, bianchi2012image, pasquini2017statistical} for forgery detection. While these traditional image manipulation detection methods are explainable and computationally efficient, most suffer from poor detection accuracy and limited generalization. 
To achieve an accurate, generalized, and interpretive image forgery localization, we introduce a learning-based framework in this work designed to capture low-level pixel inconsistency artifacts.

\subsection{Learning-based Manipulation detection and localization methods}
Recent years have witnessed significant progress in image forensics, with various learning-based methods proposed to solve the forgery localization problem, which substantially improved detection performances. Many of these methods leverage a wide range of prior knowledge, such as noise telltales~\cite{cozzolino2019noiseprint, guillaro2023trufor, zhou2018learning}, CFA artifacts~\cite{bammey2020adaptive}, and JPEG features~\cite{kwon2022learning, rao2022towards, wang2022jpeg} to perform the forgery detection. High-frequency (HF) filters~\cite{zhuo2022self, li2019localization}, such as steganalysis rich
model (SRM) filter~\cite{zhou2018learning, wu2019mantra} and Bayer filter~\cite{dong2022mvss, wu2019mantra} have also been used to capture abundant HF forgery artifacts. Besides, detecting the forgery boundary~\cite{dong2022mvss, salloum2018image} has effectively improved pixel-level forgery detection performance. In turn, some methods~\cite{dong2022mvss, liu2022pscc, gao2022generic, ferreira2016behavior} utilize multi-scale learning to extract forgery features from different levels, thereby achieving increased detection accuracy. \textcolor{black}{While SPAN \cite{hu2020span} models relationships between image patches or pixels at multiple scales using a pyramid of local self-attention blocks, our method innovatively employs a local pixel dependency encoder to capture local pixel-difference, a masked self-attention global pixel dependency encoder to model long-range pixel correlations, and feature fusion modules to combine the forgery fingerprints. These components are designed to better capture inherent pixel-inconsistency artifacts within forgery images.}
Thanks to the advent of vision transformer (ViT), ViT-based detectors~\cite{wang2022objectformer, lin2023image} take advantage of long-range interaction and no inductive bias, yielding outstanding detection performance in different problems, including forensics. However, these data-driven methods suffer from limited generalization and robustness capability.
This paper argues that pixel inconsistency within forgery images represents a more ubiquitous artifact across different manipulations and datasets. As such, we devise a novel image forgery localization framework that captures pixel inconsistency artifacts to achieve more generalized and robust forgery localization performance.

\subsection{Pixel Dependency Modeling}
Autoregressive (AR) models \cite{larochelle2011neural, germain2015made, van2016conditional, salimans2017pixelcnn++, parmar2018image, chen2018pixelsnail, child2019generating} have achieved remarkable success across various computer vision tasks, including image generation \cite{germain2015made, van2016conditional, kolesnikov2017pixelcnn}, completion \cite{chen2018pixelsnail, jain2020locally, parmar2018image}, and segmentation \cite{ouali2020autoregressive}. These AR methods aim to model the joint probability distribution of each pixel as follows:
\begin{equation}
     {\hat{a}_{i}} \sim p_{\theta}(a_{i}|a_{1},...,a_{i-1}).
\end{equation}
These models employ specific mask convolution or mask self-attention strategies, such that the probability distribution of the current pixel depends on all previous pixels in the generation order. Pioneering AR models like PixelCNN \cite{van2016conditional} and PixelRNN \cite{van2016pixel} demonstrate their effectiveness in modeling long-range pixel dependencies for natural images in the context of image generation. Follow-up variations, such as PixelCNN++ \cite{salimans2017pixelcnn++}, have been introduced to enhance image generation performance further. Furthermore, masked self-attention can also aid dependency modeling, such as image transformer \cite{parmar2018image} and sparse transformer \cite{child2019generating}. PixelSNAIL \cite{chen2018pixelsnail} combines causal convolutions with self-attention, improving image generation. Inspired by the success of pixel-dependency modeling in various generative tasks, we seek to extend upon this concept to the domain of forensic analysis. This paper introduces novel pixel-difference convolutions and masked self-attention mechanisms to capture local and global pixel inconsistency artifacts. 

\section{Proposed Method}
This section presents the proposed manipulation localization method. We first introduce the overall framework. Subsequently, we delve into the details and underlying rationales of the designed components, including the Global Pixel Dependency Modeling Module, the Local Pixel Dependency Modeling Module, and the Learning-to-Weight Module. Lastly, we introduce the proposed Pixel-Inconsistency Data Augmentation strategy and its advantages. 

\subsection{Overall Framework}
As Fig.~\ref{framework} depicts, this paper designs a two-stream image manipulation localization framework, which draws inspiration from the observation that manipulation processes, such as splicing, copy-move, and inpainting, inevitably disrupt the pixel regularity introduced by the demosaicing operation. The framework relies upon a Local Pixel Dependency Encoder and a Global Pixel Dependency Encoder to explore pixel inconsistency and context for manipulation localization. The input image is firstly split into  patches, which are then concurrently processed by the two encoders. \textcolor{black}{In the patch embedding process, we segment the input image into 4$\times$4-pixel patches. The raw pixel RGB values of each patch are flattened into a dimension of 4$\times$4$\times$3=48, and each patch token is subsequently projected to the embedding dimension. Intuitively, embedding individual RGB pixel values into single tokens can help model pixel dependency. However, this would significantly increase computational costs, as the number of tokens would equal the image's height ($H$) and width ($W$) ($H, W = 512$ in this work). To address this and achieve a reasonable tradeoff, we relax the patch size to 4$\times$4, balancing computational efficiency with the effective capture of pixel inconsistencies in manipulated images. Compared to individual pixels, the 4$\times$4-pixel token provides a more expressive representation. Our experiments show that the proposed method, with the adopted patch embedding strategy, successfully captures global and local pixel inconsistencies within manipulated images. Additionally, we adopt MLP layers in the designed transformer blocks to enhance the learning of pixel dependencies within each token \cite{tolstikhin2021mlp, lin2024mlp}.}
 
To explore long-range interaction and no inductive bias, we adopt transformer architectures as backbones of the two streams. The upper Local Pixel Dependency Encoder comprises four Difference Convolution (DC) Blocks designed to capture pixel inconsistencies in local regions. 
In turn, we introduce a Global Pixel Dependency Encoder comprising four novel masked self-attention blocks. The designed masked self-attention mechanism models global pixel dependencies within input images. Additionally, we design four Learning-to-Weight Modules (LWM) to complementarily combine global features [$f_{g1}$, $f_{g2}$, $f_{g3}$, $f_{g4}$] and local features [$f_{l1}$, $f_{l2}$, $f_{l3}$, $f_{l4}$] at multiple levels. The designed framework also incorporates a Boundary Decoder, a Forgery Decoder, and an Image Decoder. 

Notably, pixel inconsistency is most prominent in the boundary region. We, therefore, integrate the boundary auxiliary supervision to enhance the final forgery localization performance. The Forgery Decoder takes the combined features [$f_{1}$, $f_{2}$, $f_{3}$, $f_{4}$] as inputs to predict potential manipulated regions of input images, while the Image Decoder takes [$f_{g1}$, $f_{g2}$, $f_{g3}$, $f_{g4}$] as inputs and aims to reconstruct the original input image.  
Finally, we propose a novel Pixel-Inconsistency Data Augmentation (PIDA) strategy that focuses on pixel inconsistency rather than semantic forgery traces. This strategy further enhances the model's generalization and robustness capabilities. 
 

\begin{figure}[ht]
\centering
\includegraphics[scale=0.37]{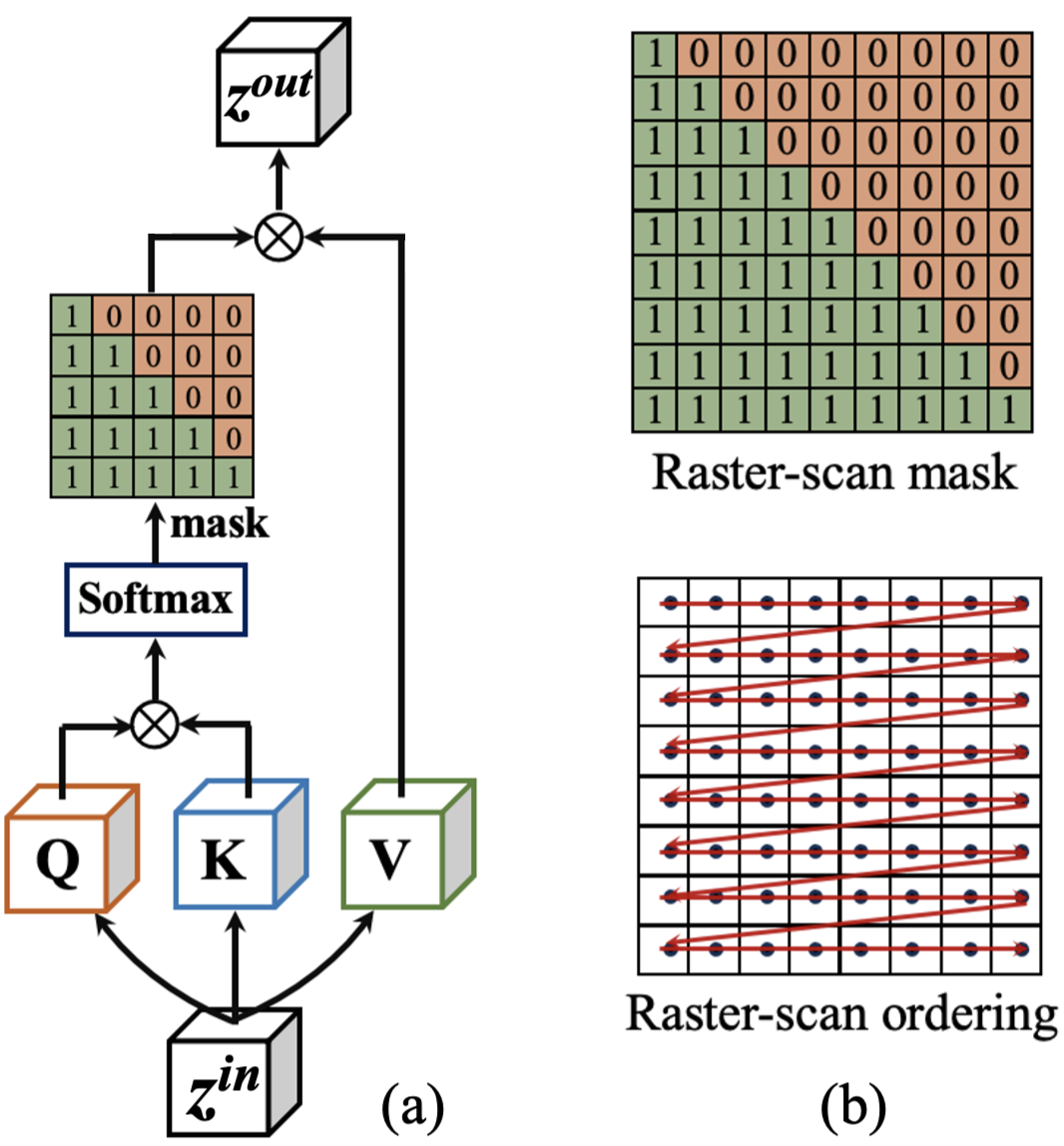}
\caption{(a). Illustration of the proposed masked attention mechanism. $\otimes$ indicates the matrix multiplication. \textbf{Q}, \textbf{K}, and \textbf{V} stand for Query, Key and Value. We designed the Raster-scan mask to model the pixel dependency. (b). The mask and corresponding pixel scan ordering. The green squares indicate the value `1' while the red squares indicate the value `0'.}
\label{maskatt}
\end{figure}

\vspace{-3mm}
\subsection{Global Pixel Dependency Modeling}
\textcolor{black}{In this part, our goal is to model the global pixel dependency across image blocks, with each token conditioned on the previous ones in a raster scan ordering. Consequently, each output token highly depends on all previous ``seen'' pixels. Compared to processing individual pixels, this design is more computational-efficient. Given the spatial redundancy in images \cite{he2022masked}, the proposed method can also effectively model global pixel dependencies. }

\textcolor{black}{Inspired by \cite{chen2018pixelsnail} and \cite{parmar2018image} that model long-term pixel dependency using attention mechanism, we introduce Masked Self-Attention (Masked SA) blocks, in a style similar to the self attention, into pixel global dependency modeling.} Fig.~\ref{framework} depicts the combination of the global pixel dependency encoder and the image decoder that forms an auto-encoder. Fig.~\ref{maskatt} (a) illustrates the details of the proposed masked self-attentions and the corresponding mask design, with \textbf{Q}, \textbf{K}, and \textbf{V} representing Query, Key, and Value, respectively. \textcolor{black}{(We omit the normalization and MLP layers for conciseness).} $z^{in}$ and $z^{out}$ indicate the input and output features. $\otimes$ denotes the matrix multiplication operator. The masked self-attention mechanism can be formulated as:  
\begin{equation}
     z^{out} = {\textbf{\rm Mask}} [{\rm softmax}(\frac{y_{query}(z^{in})y_{key}(z^{in})^\top} {\sqrt{dim}})]y_{value}(z^{in}),
\end{equation}
where $y_{query}(\cdot)$, $y_{key}(\cdot)$, and $y_{value}(\cdot)$ represent the learnable parameters, and ${y_{query}(z^{in})}$, ${y_{key}(z^{in})}$, and ${y_{value}(z^{in})}$ are equivalent to \textbf{Q}, \textbf{K}, and \textbf{V}. As Fig.~\ref{maskatt} (b) shows, we employ a raster-scan mask to model the global pixel dependency, corresponding to the raster-scan sampling ordering for the input image \cite{ouali2020autoregressive}. \textcolor{black}{If we name the input $z^{in} \in \mathbb{R}^{N \times dim} $ as $z^{in}$ =[$z^{in}_{1}$, $z^{in}_{2}$,..., $z^{in}_{N}$]$^{\top}$, then each row $z^{in}_{m}$ represents a input token. For the output $z^{out} \in \mathbb{R}^{N \times dim}$ of the proposed masked attention mechanism, each output token $z^{out}_{m}$ can be rewritten as \cite{chen2018pixelsnail}:}
\begin{equation}
     z^{out}_{m} = \sum_{n\leq m} \gamma_{mn} y_{value}(z^{in}_{n}),
\end{equation}
where elements $\gamma_{mn}$ in row $m$ can be formulated as:
\begin{footnotesize}
\begin{equation}
     \gamma_{m} = {\rm softmax}[y_{key}(z^{in}_{1})^{\top}y_{query}(z^{in}_{m}), ... ,y_{key}(z^{in}_{m})^{\top}y_{query}(z^{in}_{m})],
\end{equation}
\end{footnotesize}
\textcolor{black}{In Eq. (3), we can readily observe that each output token $z^{out}_{m}$ is  conditioned on the previous seen tokens $z^{in}_{n}$ ($n\leq m$) in the input $z^{in}$, and the scan order follows a raster-scan ordering. This mechanism also facilitates modeling more complex pixel dependencies in real-world applications, such as the dependency introduced by smart image signal processors in modern AI cameras.} As such, each conditional can access any pixel within its context through the attention operator, as indicated by the summation over all available context, denoted as $\sum_{n\leq m}$. 


\textcolor{black}{This designed module enables the access of far-away pixels, thereby enhancing the modeling of long-range statistics. As such, the extracted features [$f_{g1}$, $f_{g2}$, $f_{g3}$, $f_{g4}$] can carry abundant global pixel dependency information.} Experimental results demonstrate that the captured pixel correlations between real and manipulated images are distinctive for image forgery localization.


\subsection{Local Pixel Dependency Modeling}


According to the nature of demosaicing algorithms, the pixel correlation regularity of a given pixel largely depends on its neighboring pixels \cite{cao2009accurate, cao2009accurate2}. Moreover, the pixel regularity can be modeled by linear demosaicing formulas \cite{cao2009accurate, li2016color}. However, these traditional methods exhibit limited forgery detection performance. Inspired by \cite{yu2020searching, su2021pixel, liu2012extended}, we propose to model the local pixel dependency by integrating the traditional demosaicing ideas into convolutional operations. 

\textcolor{black}{In the Local Pixel Dependency Encoder, we place Difference Convolution (DC) heads on top of each transformer block to model pixel dependency in local image regions in a learning-based fashion. Our designed Difference Convolutions (DC) are performed at the token level, with each token representing a very small image block. Compared to processing individual pixels, the 4$\times$4 image block provides a more expressive representation for performing difference convolutions. Our method significantly reduces computational costs while effectively capturing pixel inconsistencies in local image regions. Moreover, we adopt MLP layers in each transformer block to further enhance the learning of local pixel dependencies within each block.}

\begin{figure}[ht]
\centering
\includegraphics[scale=0.24]{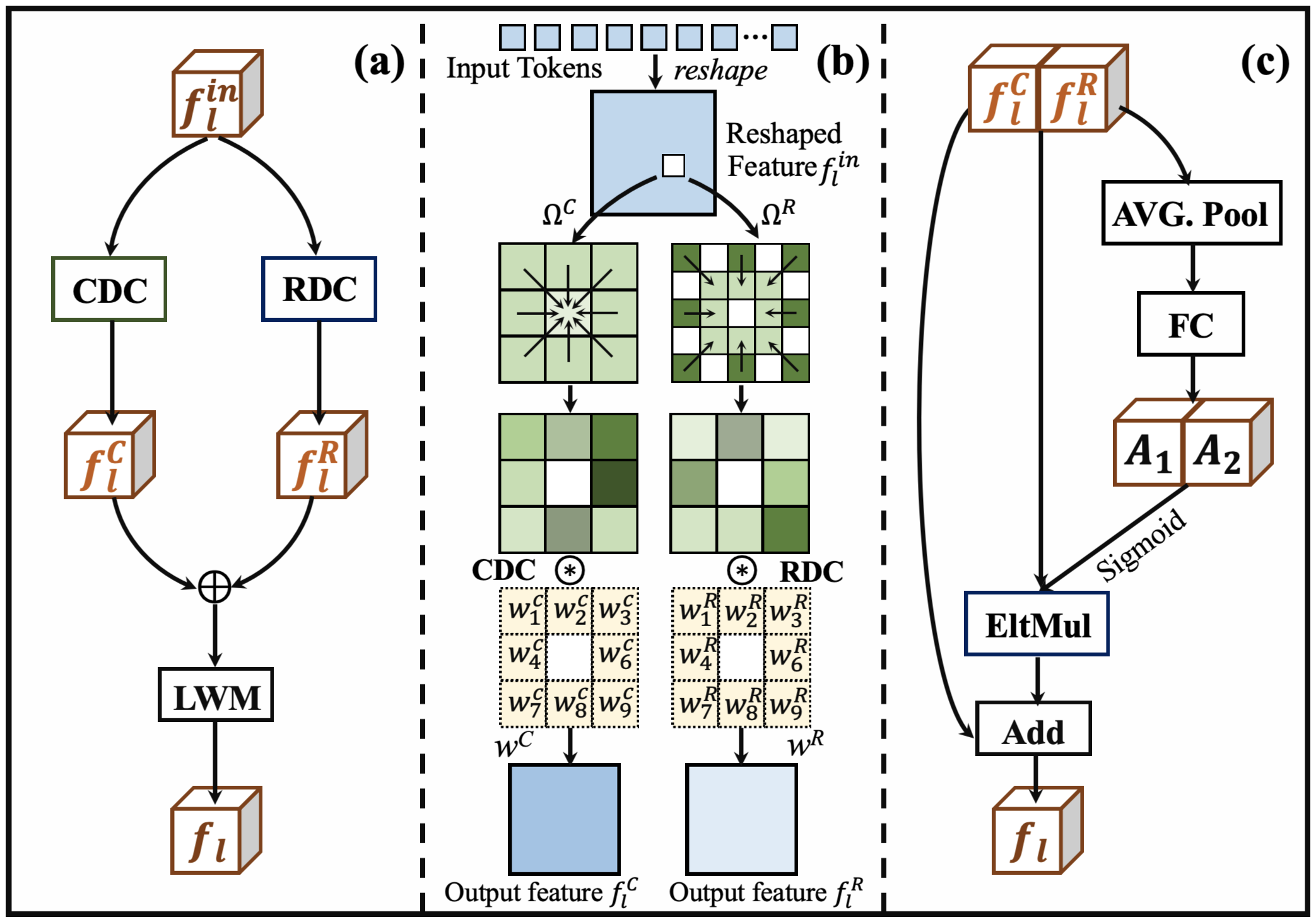}
\caption{\textcolor{black}{(a). Difference Convolution (DC) head; (b). Details of Central Difference Convolution (CDC) and Radial Difference Convolution (RDC); (c). Details of Learning-to-Weight Module (LWM).}}
\label{FFM}
\end{figure}

Fig.~\ref{FFM} (a). depicts the architecture of the designed Difference Convolution (DC) head. The input feature $f_{l}^{in}$ is firstly fed forward to two  difference convolution modules: Central Difference Convolution (CDC) and Radial Difference Convolution (RDC). By exploiting CDC and RDC, the local pixel dependencies can be effectively modeled, enhancing the final forgery localization performance. Fig.~\ref{FFM}. (b) presents the details of CDC and RDC. \textcolor{black}{The input tokens, which are the output of transformer blocks in the Local Pixel Dependency Encoder, are reshaped into a 2D feature $f_{l}^{in}$.} We first calculate the difference within local feature map regions for a given input feature map. Then, we respectively convolve the two pixel-difference feature maps with the corresponding convolutional weights, resulting in CDC and RDC feature maps. The CDC operation can be formulated as:
\begin{equation}
     f_{l}^{C} = \sum_{(x_{i},x_{c})\in {\Omega^{C}}} {w_{i}^{C}(x_{i}-x_{c})}.
\end{equation}
\textcolor{black}{Here, $x_{c}$ represents the center element in the local region $\Omega^{C}$, and $x_{i}$ denotes the corresponding surrounding elements. Each element in $\Omega^{C}$ or $\Omega^{R}$ depicted in Fig.~\ref{FFM} (b) represents a token.} The $w_{i}^{C}$ values represent learnable convolutional weights. Similarly, the RDC operation can be expressed as:
\begin{equation}
     f_{l}^{R} = \sum_{(x_{i},x_{i}^{'})\in {\Omega^{R}}} {w_{i}^{R}(x_{i}-x_{i}^{'})},
\end{equation}
where $x_{i}$ and $x_{i}^{'}$ are element pairs in region ${\Omega^{R}}$, as illustrated in Fig.~\ref{FFM} (b). 

We complementarily combine CDC features $f_{l}^{C}$ and RDC features $f_{l}^{R}$ using a Learning-to-Weight Module (LWM), which shall be elaborated in Sec. 3.4. Our designed model aims at extracting local pixel-dependency features. \textcolor{black}{Compared to the vanilla convolution, CDC and RDC benefit from their difference operations, exposing more pixel inconsistency artifacts and boosting the final image forgery localization performance.}  

\begin{figure*}[ht]
\centering
\includegraphics[scale=0.395]{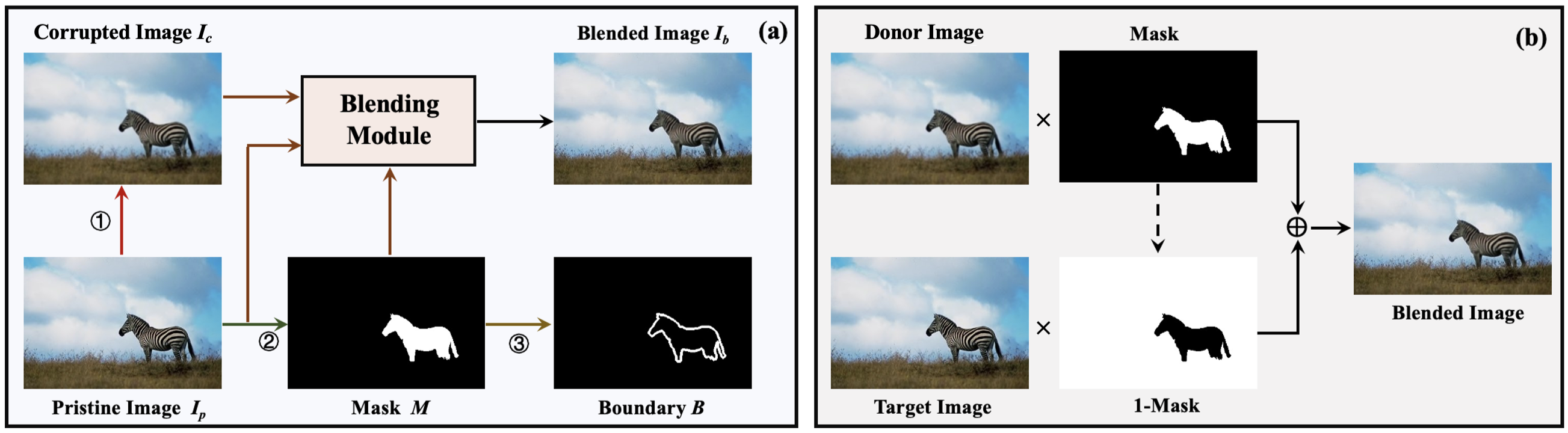}
\caption{(a). Pixel-Inconsistency Data Augmentation pipeline. \ding{192} For a given real pristine image $I_{p}$, we firstly apply common image perturbations to obtain the corrupted image $I_{c}$; \ding{193} We use built-in OpenCV function to extract the foreground mask $M$ of $I_{p}$; The Blending Module takes $I_{p}$, $I_{c}$, and $M$ as inputs and outputs the self-blended forge image $I_{b}$. \ding{194} The boundary label $B$ of the manipulated image can be obtained from $M$. (b). Details of the blending module in (a). The output blended image is the combination of the donor image's foreground and the target image's background.}
\label{Selfblending}
\end{figure*}

\subsection{Learning-to-Weight Module}
As Fig.~\ref{FFM} (a) shows, the features $f_{l}^{C}$ and $f_{l}^{R}$ generated by CDC and RDC are combined and sequentially delivered to the Learning-to-Weight Module (LWM). \textcolor{black}{The designed LWM fuses these two input features using learned weights, enabling more effective feature integration. Fig.~\ref{FFM} (c) showcases the Learning-to-Weight process for local CDC features $f_{l}^{C}$ and RDC features $f_{l}^{R}$, where \textbf{FC} and \textbf{EltMul} represent the fully-connected layer and element-wise multiplication.
In this process, the concatenated feature goes through one average pooling layer and one \textbf{FC} layer. The learned weights $A_{1}\oplus A_{2}$ are then sequentially applied to the concatenated feature $f_{l}^{C}\oplus f_{l}^{R}$ via element-wise multiplication. Finally, the fused feature $f_{l}$ is obtained by adding the concatenated feature to the weighted feature.}  

\textcolor{black}{Similarly, as depicted in Fig.~\ref{framework}, we further employ LWM to fuse the local pixel-dependency features [$f_{l1}$, $f_{l2}$, $f_{l3}$, $f_{l4}$] and the global pixel-dependency features [$f_{g1}$, $f_{g2}$, $f_{g3}$, $f_{g4}$]. The fused features [$f_{1}$, $f_{2}$, $f_{3}$, $f_{4}$] are then delivered to the boundary and forgery decoder for boundary and forgery map prediction. }

\subsection{Pixel-Inconsistency Data Augmentation}
Previous methods \cite{wang2022objectformer} mainly focus on discovering semantic-level (or object-level) inconsistencies in forgery images. Some methods \cite{dong2022mvss, zhuang2021image} also propose randomly pasting objects to pristine real images to perform data augmentation. However, as image manipulation techniques advance, forgery content's sophistication grows in tandem. Consequently, the methods designed to capture semantic-level inconsistencies struggle to generalize well to the advanced manipulations. We introduce a Pixel-Inconsistency Data Augmentation (PIDA) strategy to capture pixel-level inconsistencies instead of semantic forgery traces. Fig.~\ref{Selfblending} (a) illustrates the proposed PIDA pipeline. \ding{192} For a given real pristine image $I_{p}$, we apply image perturbations (e.g., compression, noise, and blurriness) to generate the corrupted image $I_{c}$; \ding{193} We can readily use built-in OpenCV function to extract the foreground mask $M$ of $I_{p}$. The Blending Module takes $I_{p}$, $I_{c}$, and $M$ as inputs and produces the self-blended forge image $I_{b}$; \ding{194} The boundary label $B$ of the manipulated image can be easily derived from $M$. Fig.~\ref{Selfblending} (b) details the blending module. We combine the donor image's foreground with the target image's background to generate the self-blended forgery sample.

The proposed PIDA method bears the following advantages: (1) It exclusively utilizes pristine images to generate examples of forgeries. Real data is considerably more accessible than image forgeries, facilitating training data-hungry detectors; (2) As the generated forgery samples maintain semantic consistency, the PIDA strategy directs the model's attention toward capturing pixel inconsistencies, enhancing detection performance; (3) The generated forgery samples can be regarded as harder samples, effectively increasing the difficulty of the training set. \textcolor{black}{More PIDA details can be found in the Appendix.}

\subsection{Objective Function}
The whole framework is trained in an end-to-end manner, and the overall objective function consists of the following four components: mask prediction loss $L_{M}$, boundary prediction loss $L_{B}$, compactness loss $L_{C}$, and image reconstruction loss $L_{R}$:
\begin{equation}
     L = L_{M} + {\lambda_{B}} L_{B} + {\lambda_{C}} L_{C} + {\lambda_{R}} L_{R},
\end{equation}
where $L_{M}$ and $L_{B}$ are cross-entropy losses between predicted results and the corresponding labels. The boundary loss $L_{B}$ can be considered as an auxiliary supervision for better forgery localization performance. Based on the observation that most manipulated regions are rather compact, we further apply compactness constraint $L_{C}$ to predicted masks:
\begin{small}
\begin{equation}
     L_{C} = \frac{1}{N_{img}}\sum_{{i=1}}^{N_{img}}\frac{Peri^{2}}{4\pi S} = \frac{1}{N_{img}}\sum_{{i=1}}^{N_{img}}\frac{\sum_{{j\in \hat{B}}}\hat{B_{j}}^2}{4\pi (\sum_{{k\in \hat{M}}}|\hat{M_{k}}|+\epsilon)}.
\end{equation}
\end{small}
\textcolor{black}{In this equation, $Peri$ and $S$ denote the perimeter and area of the predicted forgery region, respectively, while $N_{img}$ represents the number of images. $\hat{B}$ and $\hat{M}$ refer to the predicted boundaries and masks. As such, the nominator of Eq. (8) calculates the sum of the squared pixel values $\hat{B_{j}}^{2}$ in the predicted boundary map $\hat{B}$. The denominator is proportional to the sum of the absolute pixel values $\hat{M_{k}}$ in the predicted mask map $\hat{M}$. Here, $\epsilon$ is set to a very small value. 
Utilizing $L_{C}$ makes the predicted image forgery map more compact and improves the manipulation localization performance.} 

The image reconstruction loss $L_{R}$ calculates the $l_{1}-norm$ of the difference between the reconstructed images $\hat{I_{i}}$ and the corresponding input images $I_{i}$:
\begin{equation}
     L_{R}= \frac{1}{N}\sum_{{i=1}}^{N}
     ||I_{i}-\hat{I_{i}}||_{1}.
\end{equation}
By using $L_{R}$, the global pixel dependency can be modeled in [$f_{g1}$, $f_{g2}$, $f_{g3}$, $f_{g4}$], which is used in the LWMs for the forgery map and boundary map prediction. 




\section{Experiments and Results}
Herein, we first introduce the datasets, evaluation metrics, as well as baseline models involved in this work. Subsequently, we evaluate our model in terms of generalization and robustness under different experimental settings. We also visualize the forgery localization results to illustrate the superiority of our method. Finally, we conduct ablation studies to demonstrate the effectiveness of the designed components.

\subsection{Datasets}  
This paper adopts 12 image manipulation datasets with varying properties, images resolutions and quality. We summarize these datasets in Table~\ref{dataset}, where CM, SP, and IP denote three common image manipulation types: copy-move, splicing, and inpainting. Consistent with previous research \cite{dong2022mvss, zhou2018learning, salloum2018image}, we utilize the CASIAv2 \cite{Dong2013} dataset as the training set due to its extensive collection of over 12,000 images with diverse contents. Furthermore, we employ the DEF-12k-val \cite{mahfoudi2019defacto} as the validation set, consisting of 6,000 challenging fake images with three forgery types and 6,000 real images collected from the MS-COCO \cite{lin2014microsoft} dataset. For the testing phase, we select 11 challenging datasets, including Columbia \cite{hsu2006columbia}, IFC \cite{IFC}, CASIAv1+\footnote{CASIAv1+ and the training set CASIAv2 share 782 identical real images. To prevent data leakage, CASIAv1+ relaces these real images with the equal number of images from COREL \cite{wang2001simplicity}.} \cite{dong2013casia}, WildWeb \cite{zampoglou2015detecting}, COVER \cite{wen2016coverage}, NIST2016 \cite{guan2019mfc}, Carvalho \cite{carvalho2015illuminant}, Korus \cite{korus2016evaluation}, In-the-wild \cite{huh2018fighting}, DEF-12k-test \cite{mahfoudi2019defacto}, and IMD2020 \cite{novozamsky2020imd2020}, sorted by released dates. In all  datasets, we uniformly label forgery regions as `1' and authentic regions as `0'.

\subsection{Evaluation metrics}
This paper evaluates state-of-the-art models' pixel-level forgery detection performances using four metrics: F1, MCC, IoU, and AUC. 



\noindent\textbf{F1 Score} is a pervasive metric in binary classification, employed in image forgery detection and localization. It calculates the harmonic mean of precision and recall:   
\begin{equation}
    \label{F1}
 F1 = 2 \cdot \frac{Precision \times Recall} {Precision + Recall} = \frac{2 \times TP} {2 \times TP + FP + FN},
\end{equation}
where $TP$, $TN$, $FP$, and $FN$ represent True Positives, True Negatives, False Positives, and False Negatives.  

\noindent\textbf{Matthews Correlation Coefficient (MCC)} measures the correlation between the predicted and true values. MCC value falls within -1 and 1, where a higher MCC indicates better  performance. The calculation of MCC is derived from the formula below:  

\begin{equation}
    \label{MCC}
 MCC = \frac{TP \times TN - FP \times FN} {\sqrt{(TP+FP)(TP+FN)(TN+FP)(TN+FN)}}.
\end{equation}

\noindent\textbf{Intersection over Union (IoU)} is a widely used metric in semantic segmentation. 
The numerator of the IoU metric measures the area of intersection between prediction $P$ and ground-truth $G$, while the denominator calculates the area of the union between $P$ and $G$: 
\begin{equation}
    \label{IoU}
     IoU = \frac{P \cap G}{P \cup G}.
\end{equation}
 

\noindent\textbf{Area Under Curve (AUC)} measures the area under the Receiver Operating Characteristic (ROC) curve. Unlike the other metrics, the AUC does not require threshold selection. It quantifies the overall performance of the model across all possible thresholds. 

\begin{table}
  \centering
\captionof{table}{Summary of image manipulation datasets involved in this paper. CM, SP, and IP indicate three common image manipulation types: copy-move, splicing, and inpainting.}
\scalebox{0.71}{
\begin{tabular}{cccccccc}
    \toprule
    Dataset & Year & Venue & \#Real & \#Fake & \#CM & \#SP & \#IP \\
    \midrule
    CASIAv2\cite{Dong2013} & 2013 & ChinaSIP & 7,491 & 5,123 & 3,295 & 1,828 & 0 \\
    \midrule[1.5pt]
    DEF-12k-val\cite{mahfoudi2019defacto} & 2019 & EUSIPCO & 6,000  & 6,000 & 2,000 & 2,000 & 2,000 \\
    \midrule[1.5pt]
    Columbia\cite{hsu2006columbia} & 2006 & ICME & 183 & 180 & 0 & 180 & 0 \\
    \midrule
    IFC\cite{IFC} & 2013 & IFC-TC & 1050 & 450 & - & - & - \\
    \midrule
    CASIAv1+\cite{dong2013casia} & 2013 & ChinaSIP & 800 & 920 & 459 & 461 & 0 \\
    \midrule
    WildWeb\cite{zampoglou2015detecting} & 2015 & ICMEW & 99 & 9,657 & 0 & 9,657 & 0\\
    \midrule
    COVER\cite{wen2016coverage} & 2016 & ICIP & 100 & 100 & 100 & 0 & 0 \\
    \midrule
    NIST2016\cite{guan2019mfc} & 2016 & OpenMFC & 0 & 564 & 68 & 288 & 208 \\
    \midrule
    Carvalho\cite{carvalho2015illuminant} & 2016 & IEEE TIFS & 100 & 100 & 0 & 100 & 0 \\
    \midrule
    Korus\cite{korus2016evaluation} & 2016 & WIFS & 220 & 220 & - & - & - \\
    \midrule
    In-the-wild\cite{huh2018fighting} & 2018 & ECCV & 0 & 201 & 0 & 201 & 0 \\
    \midrule
    DEF-12k-test\cite{mahfoudi2019defacto} & 2019 & EUSIPCO & 6,000  & 6,000 & 2,000 & 2,000 & 2,000 \\
    \midrule
    IMD2020\cite{novozamsky2020imd2020} & 2020 & WACVW & 404 & 2010 & - & - & - \\
    \bottomrule
  \end{tabular}}
\label{dataset}
\end{table}

\begin{table*}
  \centering
  \caption{\textcolor{black}{Image manipulation localization performance (\textbf{F1 score} with fixed threshold: 0.5).}}
  \scalebox{0.76}{\begin{tabular}{lcccccccccccccc}
    \toprule
    Method & Venue & NIST & Columbia & CASIAv1+ & COVER & DEF-12k & IMD & Carvalho & IFC & In-the-Wild & Korus & WildWeb & AVG \\
    \midrule
    FCN \cite{long2015fully} & CVPR15 & .167 & .223 & .441 & .199 & .130 & .210 & .068 & .079 & .192 & .122 & .110 & .176 \\
    \midrule 	
    U-Net \cite{ronneberger2015u} & MICCAI15 & .173 & .152 & .249 & .107 & .045 & .148 & .124 & .070 & .175 & .117 & .056 & .129\\
    \midrule
    DeepLabv3 \cite{chen2017deeplab} & TPAMI18 & .237 & .442 & .429 & .151 & .068 & .216 & .164 & .081 & .220 & {.120} & .098 & .202\\ 
    \midrule
    MFCN \cite{salloum2018image} & JVCIP18 & .243 & .184 & .346 & .148 & .067 & .170 & .150 & .098 & .161 & .118 & .102 & .162 \\ 
    \midrule 	
    RRU-Net \cite{bi2019rru} & CVPRW19 & .200 & .264 & .291 & .078 & .033 & .159 & .084 & .052 & .178 & .097 & .092 & .139  \\ 
    \midrule
    MantraNet \cite{wu2019mantra} & CVPR19 & .158 & .452 & .187 & .236 & .067 & .164 & .255 & {.117} & {.314} & .110 & \underline{.224} & .208\\ 
    \midrule
    HPFCN \cite{li2019localization} & ICCV19 & .172 & .115 & .173 & .104 & .038 & .111 & .082 & .065 & .125 & .097 & .075 & .105\\ 
    \midrule
    H-LSTM \cite{bappy2019hybrid} & TIP19 & \textbf{.357} & .149 & .156 & .163 & .059 & .202 & .142 & .074 & .173 & .143 & .141 & .160\\ 
    \midrule
    SPAN \cite{hu2020span} & ECCV20 & .211 & .503 & .143 & .144 & .036 & .145 & .082 & .056 & .196 & .086 & .024 & .148 \\
    \midrule
    ViT-B \cite{dosovitskiy2020vit} & ICLR21 & .254 & .217 & .282 & .142 & .062 & .154 & .169 & .071 & .208 & \underline{.176} & .117 & .168\\ 
    \midrule
    Swin-ViT \cite{liu2021swin} & ICCV21 & .220 & .365 & .390 & .168 & {.157} & {.300} & .183 & .102 & .265 & .134 & .040 & .211 \\
    \midrule
    PSCC \cite{liu2022pscc} & TCSVT22 & .173 & .503 & .335 & .220 & .072 & .197 & \textbf{.295} & .114 & .303 & .114 & .112 & .222 \\
    \midrule
    MVSS-Net++ \cite{dong2022mvss} & TPAMI22 & \underline{.304} & {.660} & {.513} & \textbf{.482} & .095 & .270 & \underline{.271} & .080 & .295 & .102 & .047 & {.284} \\	
    \midrule		
    CAT-NET \cite{kwon2022learning} & IJCV22 & .102 & .206 & .237 & .210 & \textbf{.206} & .257 & .175 & .099 & .217 & .085 & .170 & .179\\
    \midrule
    EVP \cite{liu2023evp} & CVPR23 & .210 & .277 & .483 & .114 & .090 & .233 & .060 & .081 & .231 & .113 & .099 & .181 \\
    \midrule
    \textcolor{black}{TruFor} \cite{guillaro2023trufor} & CVPR23 & .268 & \textbf{.829} & \underline{.532} & \underline{.280} & .148 & \underline{.359} & .213 & \underline{.127} & \underline{.361} & .122 & .169 & \underline{.310} \\ 
    \midrule[1.5pt]
    \cellcolor[HTML]{E0DBDB}PIM & \cellcolor[HTML]{E0DBDB}Ours & \cellcolor[HTML]{E0DBDB}.280 & \cellcolor[HTML]{E0DBDB}\underline{.680} & \cellcolor[HTML]{E0DBDB}\textbf{.566} & \cellcolor[HTML]{E0DBDB}.251 & \cellcolor[HTML]{E0DBDB}\underline{.167} & \cellcolor[HTML]{E0DBDB}\textbf{.419} & \cellcolor[HTML]{E0DBDB}.253 & \cellcolor[HTML]{E0DBDB}\textbf{.155} & \cellcolor[HTML]{E0DBDB}\textbf{.418} & \cellcolor[HTML]{E0DBDB}\textbf{.234} & \cellcolor[HTML]{E0DBDB}\textbf{.236} & \cellcolor[HTML]{E0DBDB}\textbf{.333}\\
    \bottomrule
  \end{tabular}}
  \label{f1_table}
\end{table*}

\begin{table*}
  \centering
  \caption{\textcolor{black}{Image manipulation localization performance (\textbf{IoU score} with fixed threshold: 0.5).}}
  \scalebox{0.76}{\begin{tabular}{lcccccccccccccc}
    \toprule
    Method & Venue & NIST & Columbia & CASIAv1+ & COVER & DEF-12k & IMD & Carvalho & IFC & In-the-Wild & Korus & WildWeb & AVG \\
    \midrule
    FCN \cite{long2015fully} & CVPR15 & .114 & .177 & .367 & .117 & .089 & .158 & .043 & .058 & .140 & .089 & .084 & .131\\ 		
    \midrule 
    U-Net \cite{ronneberger2015u} & MICCAI15 & .128 & .097 & .204 & .072 & .031 & .105 & .082 & .048 & .121 & .082 & .044 & .092 \\ 		
    \midrule 
    DeepLabv3 \cite{chen2017deeplab} & TPAMI18 & .191 & .353 & .361 & .106 & .050 & .159 & .112 & .058 & .162 & .084 & .073 & .155\\ 			
    \midrule
    MFCN \cite{salloum2018image} & JVCIP18 & .193 & .123 & .291 & .100 & .050 & .124 & .103 & .074 & .112 & .083 & .080 & .121 \\ 
    \midrule 		
    RRU-Net \cite{bi2019rru} & CVPRW19 & .156 & .196 & .244 & .057 & .024 & .119 & .057 & .039 & .131 & .068 & .080 &  .106 \\ 
    \midrule 											
    MantraNet \cite{wu2019mantra} & CVPR19 & .098 & .301 & .111 & .139 & .039 & .098 & .153 & .068 & .201 & .061 & \underline{.146} & .129\\  
    \midrule 
    HPFCN \cite{li2019localization} & ICCV19 & .126 & .076 & .137 & .070 & .026 & .076 & .054 & .045 & .084 & .064 & .057 & .074\\	
    \midrule	
    H-LSTM \cite{bappy2019hybrid} & TIP19 & \textbf{.276} & .090 & .101 & .108 & .037 & .131 & .084 & .047 & .106 & .094 & .095 & .106\\ 	
    \midrule
    SPAN \cite{hu2020span} & ECCV20 & .156 & .390 & .112 & .105 & .024 & .100 & .049 & .037 & .132 & .055 & .015 & .107\\ 	
    \midrule
    ViT-B \cite{dosovitskiy2020vit} & ICLR21 & .197 & .164 & .232 & .101 & .045 & .192 & .121 & .051 & .152 & \underline{.130} & .094 & .134\\
    \midrule	
    Swin-ViT \cite{liu2021swin} & ICCV21&  .167 & .297 & .356 & .124 & {.129} & {.243} & .132 & {.078} & .214 & .103 & .033 & .171\\
    \midrule
    PSCC \cite{liu2022pscc} & TCSVT22 & .108 & .360 & .232 & .130 & .042 & .120 & {.185} & .067 & .193 & .066 & .070 & .143\\  								
    \midrule										
    MVSS-Net++ \cite{dong2022mvss} & TPAMI22 & \underline{.239} & {.573} & {.397} & \textbf{.384} & .076 & .200 & \underline{.188} & .055 & {.219} & .075 & .034 & {.222}\\  
    \midrule	
    CAT-NET \cite{kwon2022learning} & IJCV22 & .062 & .140 & .165 & .141 & \textbf{.152} & .183 & .110 & .062 & .144 & .049 & .107 & .120\\		
    \midrule 											
    EVP \cite{liu2023evp} & CVPR23 & .160 & .213 & {.421} & .083 & .070 & .183 & .043 & .062 & .182 & .084 & .071 & .143 \\	
     \midrule 
    \textcolor{black}{TruFor} \cite{guillaro2023trufor} & CVPR23 & .212 & \textbf{.781} & \underline{.481} & \underline{.215} & .121 & \underline{.297} & .159 & \underline{.100} & \underline{.303} & .095 & .138 & \underline{.264} \\
    \midrule[1.5pt]
    \cellcolor[HTML]{E0DBDB}PIM & \cellcolor[HTML]{E0DBDB}Ours & \cellcolor[HTML]{E0DBDB}.225 & \cellcolor[HTML]{E0DBDB}\underline{.604} & \cellcolor[HTML]{E0DBDB}\textbf{.512} & \cellcolor[HTML]{E0DBDB}{.188} & \cellcolor[HTML]{E0DBDB}\underline{.133} & \cellcolor[HTML]{E0DBDB}\textbf{.340} & \cellcolor[HTML]{E0DBDB}\textbf{.194} & \cellcolor[HTML]{E0DBDB}\textbf{.119} & \cellcolor[HTML]{E0DBDB}\textbf{.338} & \cellcolor[HTML]{E0DBDB}\textbf{.182} & \cellcolor[HTML]{E0DBDB}\textbf{.193} & \cellcolor[HTML]{E0DBDB}\textbf{.275} \\
    \bottomrule	
  \end{tabular}}
  \label{iou_table}
\end{table*}

\subsection{Baseline Models} 
This paper incorporates 16 representative baseline detectors from top journals and conferences, including five data-driven architectures and 11 state-of-the-art image forgery detectors. The goal is to evaluate the detection performance of different network architectures and facilitate a head-to-head comparison. The baselines include three pervasive CNN architectures (FCN \cite{long2015fully}, U-Net \cite{ronneberger2015u}, and DeepLabv3 \cite{chen2017deeplab}) and two vision transformers (ViT-B \cite{dosovitskiy2020vit} and Swin-ViT \cite{liu2021swin}). Furthermore, this benchmark incorporates ten state-of-the-art image forgery detection models: 

\noindent\textbf{MFCN} \cite{salloum2018image} casts the image splicing localization as a multi-task problem. It exploits the two-branch FCN VGG-16 network to predict the forgery map and boundary map simultaneously.

\noindent\textbf{RRU-Net} \cite{bi2019rru} is an end-to-end ringed residual U-Net architecture specifically designed for image splicing detection. It leverages residual propagation to address the issue of gradient perturbation in deep networks effectively. By incorporating this mechanism, RRU-Net strengthens the learning process of forgery clues.

\noindent\textbf{MantraNet} \cite{wu2019mantra} is an end-to-end image forgery detection and localization framework trained on a dataset consisting of 385 manipulation types. To achieve robust image manipulation detection, MantraNet introduces a novel long short-term memory solution specifically designed to detect local anomalies.

\noindent\textbf{HPFCN} \cite{li2019localization} ensembles the ResNet blocks and a learnable high-pass filter to perform the pixel-wise inpainting localization.  

\noindent\textbf{H-LSTM} \cite{bappy2019hybrid} is a forgery detection model that integrates both a CNN encoder and LSTM networks. This combination enables the model to capture and analyze spatial and frequency domain artifacts in forgery images.

\noindent\textbf{SPAN} \cite{hu2020span} is a framework that constructs a pyramid attention network to capture the interdependencies between image patches across multiple scales. It builds upon the foundation of the pre-trained MantraNet and offers the flexibility to fine-tune its parameters on specific training sets.

\noindent\textbf{PSCC} \cite{liu2022pscc} is a progressive spatial-channel correlation network, which extracts local and global features at multiple scales with dense cross-connections. The progressive learning mechanism enables the model to predict the forgery mask in a coarse-to-fine manner, thereby empowering the final detection performance. 

\noindent\textbf{MVSS-Net++} \cite{dong2022mvss} designs a two-stream network to capture boundary and noise artifacts using multi-scale features. Incorporating two streams effectively analyzes different aspects of the image to detect manipulations at both pixel and image levels.  

\noindent\textbf{CAT-NET} \cite{kwon2022learning} is a CNN-based model that leverages discrete cosine transform (DCT) coefficients to capture JPEG compression artifacts in manipulated images. 

\noindent\textbf{EVP} \cite{liu2023evp} presents a unified low-level structure detection framework for images. ViT Adaptors and visual promptings enable the EVP model to achieve outstanding forgery localization accuracy. 

\noindent \textcolor{black}{\textbf{TruFor} \footnote{For a head-to-head comparison, we align the TruFor training, validation, and testing sets with ours. } \cite{guillaro2023trufor} concurrently captures high-level RGB artifacts and low-level noise forgery traces through a transformer-based fusion architecture based on a learned noise-sensitive fingerprint.} 

\textcolor{black}{In this work, for a fair and reproducible comparison, we follow MVSS-Net++ \cite{dong2022mvss}, selecting baseline models that meet one of the following three criteria: (1) official training code is publicly available; (2) the model uses the same training protocol as ours, i.e., CASIAv2 is used as the training dataset; or (3) official pretrained models are released. During testing, we follow the protocols of MVSS-Net++ \cite{dong2022mvss} and JPEG-SSDA \cite{rao2022towards}, testing the trained models on forgery images and reporting the image-level detection results for all testing datasets.} The selected manipulation methods encompass a wide variety of forgery fingerprints, such as boundary artifacts (MFCN \cite{salloum2018image}, MVSS-Net++ \cite{dong2022mvss}), multi-scale features (PSCC \cite{liu2022pscc}, MVSS-Net++ \cite{dong2022mvss}, \textcolor{black}{TruFor \cite{guillaro2023trufor}}), high-frequency artifacts (HPFCN \cite{li2019localization}, MVSS-Net++ \cite{dong2022mvss}, MantraNet \cite{wu2019mantra}), and compression artifacts (CAT-NET \cite{kwon2022learning}). 


\subsection{Implementation Details} 
Our models are implemented in PyTorch \cite{paszke2019pytorch} and trained on two Quadro RTX 8000 GPUs. The input image size is 512 $\times$ 512. We use Adam optimizer \cite{kingma2014adam} with $\beta_{1}$=0.9 and $\beta_{2}$=0.999 to train the designed model with batch size 28. The learning rate and weight decay are 6e-5 and 1e-5, respectively. The model is trained for 20 epochs and validated every 1,600 global steps. Following the experimental setting of \cite{dong2022mvss}, we train our model on CASIAv2 \cite{Dong2013} dataset and validate it on DEF-12k-val \cite{mahfoudi2019defacto} dataset. Besides the proposed Pixel-Inconsistency Data Augmentation, we follow \cite{dong2022mvss} to use common data augmentation for training, including flipping, blurriness, compression, noise, pasting, and inpainting.


\begin{table*}
  \centering
  \caption{\textcolor{black}{Image-level manipulation detection performance (F1 score with fixed threshold: 0.5).}}
  \scalebox{0.76}{\begin{tabular}{lcccccccccccccc}
    \toprule
    Method & Venue &  NIST & Columbia & CASIAv1+ & COVER & DEF-12k & IMD & Carvalho & IFC & In-the-Wild & Korus & WildWeb & AVG \\
    \midrule
    FCN \cite{long2015fully} & CVPR15 & .897 & .702 & .713 & .653 & .607 & .827 & .566 & .441 & .908 & .627 & .769 & .701 \\ \midrule 	
    U-Net \cite{ronneberger2015u} & MICCAI15 & .945 & .692 & .673 & \underline{.660} & .633 & .878 & .662 & \underline{.466} & .972 & .637 & .715 & .721 \\ \midrule 		
    DeepLabv3 \cite{chen2017deeplab} & TPAMI18 & .939 & {.724} & .746 &  \underline{.660} & .626 & .867 & .646 & .441 & .974 & .610 & .827 & .733 \\ \midrule												
    RRU-Net \cite{bi2019rru} & CVPRW19 & .871 & .678 & .661 & .553 & .564 & .798 & .646 & .387 & .877 & .587 & .602 & .657 \\ \midrule	
    HPFCN \cite{li2019localization} & ICCV19 & .893 & .664 & .580 & .624 & .615 & .824 & .636 & .446 & .902 & .632 & .715 & .685 \\ \midrule 	
    ViT-B \cite{dosovitskiy2020vit} & ICLR21 & .969 & .707 & .653 & \textbf{.671} & \underline{.646}  & .870  & .664  & .448 & .972 & .644 & .829 & \underline{.734} \\ \midrule 	
    PSCC \cite{liu2022pscc} & TCSVT22 & .953 & .698  & .577  & \underline{.660}  & \underline{.646}  & .866& \textbf{.674} & .463  & .972 & .649 & .812 & .725 \\ \midrule	
    MVSS-Net++ \cite{dong2022mvss} & TPAMI22 & .831 & \underline{.735} & \underline{.758} & .659 & \underline{.646} & .863 & .613 & \textbf{.472} & .953 & .613 & .540 & .698 \\	
    \midrule				
    CAT-NET \cite{kwon2022learning} & IJCV22 & \textbf{.982} & .687 & .548 & .641 & .642 & \underline{.885} & .662 & .464  & \textbf{.992} & \textbf{.668} & .685 & .714 \\	\midrule 							
    EVP \cite{liu2023evp} & CVPR23 & .878 & .623 & .746  & .569 & .563 & .813 & .554 &  .418 & .828  & .573 & \underline{.888} & .678 \\	
    \midrule
    \textcolor{black}{TruFor} \cite{guillaro2023trufor} & CVPR23 & .858 & \textbf{.740} & .743  & .643 & .569 & .821 & .610 &  .414 & .886  & .530 & .760 & .689 \\
    \midrule[1.5pt] 				
    \cellcolor[HTML]{E0DBDB}PIM & \cellcolor[HTML]{E0DBDB}Ours & \cellcolor[HTML]{E0DBDB}\underline{.973}  & \cellcolor[HTML]{E0DBDB}.702 & \cellcolor[HTML]{E0DBDB}\textbf{.779}  & \cellcolor[HTML]{E0DBDB}.655  & \cellcolor[HTML]{E0DBDB}\textbf{.651} & \cellcolor[HTML]{E0DBDB}\textbf{.896}  & \cellcolor[HTML]{E0DBDB}\underline{.669}  & \cellcolor[HTML]{E0DBDB}.458  & \cellcolor[HTML]{E0DBDB}\underline{.977}  & \cellcolor[HTML]{E0DBDB}\underline{.657} & \cellcolor[HTML]{E0DBDB}\textbf{.932} & \cellcolor[HTML]{E0DBDB}\textbf{.759} \\
    \bottomrule				
  \end{tabular}}
  \label{f1_img}
\end{table*}

\subsection{Cross-Dataset Evaluation} 
\noindent\textbf{Pixel-level evaluation.} Localizing manipulated regions in forgery images is crucial as it provides evidence regarding the regions that have been manipulated. Predicted forgery regions can unveil the potential intents of attackers \cite{kong2022detect}. However, most detectors suffer from poor localization performance in cross-dataset evaluations due to substantial domain gaps between the training and testing sets. Herein, we evaluate the generalization capability of different detectors in terms of pixel-level forgery detection (i.e., manipulation localization).
In line with the cross-dataset evaluation protocols in \cite{dong2022mvss}, we train our model on CASIAv2 \cite{Dong2013} dataset and validate it on DEF-12k-val \cite{mahfoudi2019defacto} dataset. To facilitate a comprehensive interpretation of the results, we report two key metrics, namely F1 and IoU, in Table~\ref{f1_table} and Table~\ref{iou_table}, which have been widely used in image forgery localization. We further provide the AUC and MCC results in the Appendix. We highlight the best localization results in bold and underline the second-best results. Unlike in \cite{dong2022mvss} where optimal thresholds are determined individually for each model and dataset, we set the default decision threshold of F1, MCC, and IoU as 0.5 for the following two reasons: (1). In real-world application scenarios, it is unlikely to predefine different optimal threshold values for each testing data sample, and (2). Unifying the decision threshold as 0.5 enables us to compare all baseline models fairly. The pixel-level evaluation at different thresholds is presented in the Appendix.

F1-score is the most widely used metric in this field \cite{ying2023learning, qi2022principled, rao2022towards, dong2013casia}. In Table~\ref{f1_table}, our method achieves the best detection F1-score on six datasets and the second-best performance on two datasets. \textcolor{black}{ In comparison to the state-of-the-art method TruFor \cite{guillaro2023trufor}, the proposed Pixel-Inconsistency Modelling (PIM) method demonstrates  superior forgery localization F1-scores across nine datasets, with an average improvement of 2.3$\%$ average F1-score improvement, increasing from 31.0$\%$ to 33.3$\%$.}
 In Table~\ref{iou_table}, our method Pixel-Inconsistency Modelling (PIM) consistently achieves the best or second-best detection performance on unseen testing datasets. Even though the 11 unseen datasets exhibit diverse distributions, our method's average IoU score outperform all previous approaches by a significant margin. The superiority of the proposed method can be attributed to its ability to capture pixel inconsistency artifacts, which serve as a common fingerprint across different forgery datasets. 



\noindent\textbf{Image-level evaluation.}
In this subsection, we further evaluate the image-level forgery detection under cross-dataset evaluation. Ideally, the tampering probability map should all be zero for a pristine real image. To this end, we employ maximum pooling on the tampering probability map and utilize the resulting output score as the overall prediction for the input image \cite{rao2022towards}. We present the key metric F1 score in Table \ref{f1_img}. We highlight the best results in bold and underline the second-best results. Notably, our method achieves the top-2 image-level detection performance on eight datasets: NIST, CASIAv1+, DEF-12k, IMD, Carvalho, In-the-Wild, Korus, and WildWeb. Even in cases where our method ranks 6th on the COVER dataset and 5th on the IFC datasets, it closely approaches the best detection results (\textbf{COVER}: \underline{Ours: .655} v.s. \underline{Best: .671}; \textbf{IFC}: \underline{Ours: .458} v.s. \underline{Best: .472}). Our method achieves the best average results, demonstrating its outstanding forgery detection generalization performance. 


\vspace{-6mm}
\textcolor{black}{\subsection{Cross-Manipulation Evaluation}} 

\textcolor{black}{To evaluate the model's generalization capability to unseen manipulation techniques, we train our model on the CASIAv2 dataset and test it on the unseen Inpainting (IP) manipulation. The cross-manipulation F1 score on 10 inpainting techniques is presented in Table~\ref{f1_ip}, and the IoU performance can be found in the Appendix. The 10 typical and challenging inpainting datasets include CA \cite{yu2018generative}, EC \cite{nazeri2019edgeconnect}, GC \cite{yu2019free}, LB \cite{wu2021deep}, LR \cite{guo2017patch}, NS \cite{bertalmio2001navier}, PM \cite{herling2014high}, RN \cite{yu2020region}, SG \cite{huang2014image}, SH \cite{yan2018shift}, and TE \cite{telea2004image}, which are widely used in previous inpainting detection works \cite{wu2021iid}. From Table~\ref{f1_ip}, it can be observed that CAT-NET and TruFor benefit from their extensive training data and their ability to capture low-level artifacts, achieving promising average forgery localization performance. However, our proposed method PIM achieves the highest F1 scores on eight inpainting datasets, with F1 score of 0.649 on average, outperforming previous methods by a significant margin.}

\textcolor{black}{Our method’s superior generalizability to unforeseen manipulation techniques can be attributed to two key designs: (1) The Pixel-Inconsistency Data Augmentation (PIDA) strategy enables the model to capture more general and subtle artifacts, effectively mitigating overfitting during training; (2) The designed network effectively captures both global and local pixel inconsistency artifacts, enabling the model to reveal more inherent pixel-level artifacts rather than semantic traces.}

\begin{table*}
  \centering
  \caption{\textcolor{black}{Image manipulation localization performance (\textbf{F1 score} with fixed threshold: 0.5) on the unseen manipulation type: Inpainting.}}
  \scalebox{0.80}{\begin{tabular}{lccccccccccccc}
    \toprule 	
    Method & Venue & CA & EC & GC & LB & LR & NS & PM & RN & SG & SH & TE & AVG \\
    \midrule 
    FCN \cite{long2015fully} & CVPR15 & .089 & .032 & .009 & .026 & .468 & .136 & .230 & .120 & .304 & .106 & .063 & .144 \\
    \midrule 	
    U-Net \cite{ronneberger2015u} & MICCAI15 & .010 & .011 & .007 & .004 & .334 & .543 & .104 & .060 & .066 & .044 & .507 & .154 \\
    \midrule 		
    DeepLabv3 \cite{chen2017deeplab} & TPAMI18 & .105 & .069 & .011 & .021 & .566 & .648 & .265 & .185 & .467 & .131 & .593 & .278 \\
    \midrule 	
    MFCN \cite{salloum2018image} & JVCIP18 & .012 & .018 & .003 & .011 & .169 & .588 & .044 & .059 & .042 & .057 & .574 & .143 \\
    \midrule 										
    RRU-Net \cite{bi2019rru} & CVPRW19 & .036 & .054 & .029 & .021 & .452 & .538 & .194 & .096 & .177 & .078 & .444 & .193 \\
    \midrule 											
    MantraNet \cite{wu2019mantra} & CVPR19 & .270 & .419 & \underline{.272} & .395 & .070 & .425 & .045 & .294 & .107 & .355 & .354 & .273 \\
    \midrule 										
    HPFCN \cite{li2019localization} & ICCV19 & .011 & .012 & .008 & .008 & .154 & .490 & .020 & .035 & .017 & .030 & .447 & .112 \\
    \midrule 										
    H-LSTM \cite{bappy2019hybrid} & TIP19 & .049 & .033 & .043 & .039 & .117 & .059 & .043 & .062 & .038 & .048 & .049 & .053 \\
    \midrule 
    SPAN \cite{hu2020span} & ECCV20 & .009 & .031 & .009 & .005 & .357 & .432 & .116 & .108 & .184 & .017 & .224 & .136 \\
    \midrule 										
    ViT-B \cite{dosovitskiy2020vit} & ICLR21 & .021 & .018 & .016 & .029 & .103 & .354 & .020 & .035 & .030 & .049 & .339 & .092 \\
    \midrule 										
    Swin-ViT \cite{liu2021swin} & ICCV21 & .206 & .221 & .005 & .071 & .377 & .218 & \underline{.402} & .296 & .335 & .266 & .064 & .224 \\
    \midrule 										
    PSCC \cite{liu2022pscc} & TCSVT22 & .314 & .314 & .108 & .201 & .292 & .652 & .191 & .279 & .349 & .238 & .613 & .323 \\
    \midrule 										
    MVSS-Net++ \cite{dong2022mvss} & TPAMI22 & .087 & .049 & .012 & .020 & \underline{.575} & \underline{.814} & .313 & .233 & .390 & .192 & \underline{.809} & .318 \\
    \midrule 											
    CAT-NET \cite{kwon2022learning} & IJCV22 & \underline{.547} & \underline{.530} & \textbf{.382} & \underline{.757} & .335 & .459 & .244 & \textbf{.550} & \underline{.572} & \underline{.623} & .469 & \underline{.497}\\
    \midrule 
    EVP \cite{liu2023evp} & CVPR23 & .277 & .375 & .058 & .398 & .484 & .312 & .350 & .340 & .499 & .534 & .300 & .357 \\
    \midrule 											
    TruFor \cite{guillaro2023trufor} & CVPR23 & .181 & .158 & .166 & .301 & .162 & .200 & .066 & .145 & .104 & .123 & .199 & .164 \\
    \midrule[1.5pt] 
    \cellcolor[HTML]{E0DBDB}PIM & \cellcolor[HTML]{E0DBDB}Ours & \cellcolor[HTML]{E0DBDB}\textbf{.628} & \cellcolor[HTML]{E0DBDB}\textbf{.660} & \cellcolor[HTML]{E0DBDB}.080 & \cellcolor[HTML]{E0DBDB}\textbf{.790} & \cellcolor[HTML]{E0DBDB}\textbf{.774} & \cellcolor[HTML]{E0DBDB}\textbf{.836} & \cellcolor[HTML]{E0DBDB}\textbf{.537} & \cellcolor[HTML]{E0DBDB}\underline{.457} & \cellcolor[HTML]{E0DBDB}\textbf{.890} & \cellcolor[HTML]{E0DBDB}\textbf{.631} & \cellcolor[HTML]{E0DBDB}\textbf{.853} & \cellcolor[HTML]{E0DBDB}\textbf{.649}\\
    \bottomrule										
  \end{tabular}}
  \label{f1_ip}
\end{table*}

\vspace{-2mm}
\textcolor{black}{\subsection{Generalization to Sophisticated Manipulations}} 
\textcolor{black}{To examine our model's generalizability to sophisticated manipulations, we test our model on two datasets: Dall-E2 (DE2) and Stable Diffusion (SD). DE2 and SD include 60 and 328 sophisticated fake images, respectively. The forgery images exhibit high-level harmonization, with the forgery regions having compatible illumination, reasonable size, semantic consistency, and appropriate position. The generation pipelines of the two sophisticated datasets are detailed in the Appendix. The forgery localization performances (F1, IoU, AUC, and MCC scores) on unseen sophisticated manipulation techniques are shown in Table~\ref{Sophisticated}. While the state-of-the-art TruFor achieves decent localization performance in terms of the listed metrics, our proposed method, Pixel-Inconsistency Modelling (PIM), outperforms all other methods across most metrics. For both DE2 and SD datasets, PIM achieves the highest F1, IoU, and MCC scores, indicating superior generalization capability in image forgery localization for sophisticated manipulations. The superiority of PIM on sophisticated manipulations generated by advanced AIGC technologies suggests that PIM is highly effective at generalizing to unseen and complex manipulation techniques, ensuring a robust model for real-world applications.} 

\vspace{-2mm}
\textcolor{black}{\subsection{Generalization to Advanced Manipulations}} 
\textcolor{black}{With the rapid development of AIGC technologies, forgery images are becoming increasingly photorealistic, and the barrier to using AIGC tools is much lower. Therefore, it is crucial to detect these emerging advanced manipulations. We adapt our model to two image manipulation datasets: Autosplice \cite{jia2023autosplice} and CocoGlide \cite{nichol2021glide}, which are generated by advanced AIGC methodologies. Autosplice \cite{jia2023autosplice} is a text-prompt manipulated image dataset generated by powerful large vision language models. It includes 2,273 real images and 3,621 manipulated images, with each forgery image having three JPEG compression quality factors: 75, 90, and 100 (with higher values indicating better image quality). CocoGlide includes 512 photorealistic forgery images, generated from the COCO 2017 validation set using the text-guided GLIDE diffusion model. The image forgery localization scores (AUC and MCC) are reported in Table~\ref{advanced}. Our method PIM consistently achieves the best AUC and MCC performances across the Autosplice 100, Autosplice 90, and CocoGlide datasets. The SOTA method, TruFor, benefits from its Noiseprint++ extractor trained on extensive extra data, achieving the highest scores on the low-quality Autosplice 75 dataset. Nonetheless, PIM exhibits superior average AUC and MCC across all advanced AIGC datasets.}

\begin{table}[]
  \caption{\textcolor{black}{Image manipulation localization performance on unseen sophisticated manipulations. (DE2: Dall-E2, SD: Stable Diffusion)}}
  \label{Sophisticated}
\centering
\scalebox{0.80}{\begin{tabular}{lcccccccc}
\toprule
\multirow{2}{*}{Method} & \multicolumn{2}{c}{F1} & \multicolumn{2}{c}{IoU} & \multicolumn{2}{c}{AUC} & \multicolumn{2}{c}{MCC} \\ \cline{2-9} 
& DE2 & SD & DE2 & SD & DE2 & SD & DE2 & SD \\ \midrule
 FCN \cite{long2015fully}  & .122 & .248 & .065 & .141 & .708 & .847 & .137 & .250 \\
\midrule 
 U-Net  \cite{ronneberger2015u} & .314 & .173 & \underline{.186} & .095 & .921 & .834 & \underline{.314} & .170 \\
\midrule 
DeepLabv3  \cite{chen2017deeplab} & .116 & .171 & .062 & .094 & .825 & .807 & .110 & .166 \\
\midrule  
 MFCN \cite{salloum2018image} & .178 & .171 & .097 & .093 & .806 & .692 & .180  & .166 \\
\midrule 
 RRU-Net \cite{bi2019rru}   & .253 & .118 & .145 & .063 & .922 & .802 & .262 & .113 \\
\midrule  
 MantraNet \cite{wu2019mantra}  & .021 & .012 & .011 & .006 & .839 & .770 & .000 & .000 \\
\midrule  
 HPFCN  \cite{li2019localization} & .122 & .087 & .065 & .045 & .831 & .694 & .112  & .082 \\
\midrule  
  H-LSTM \cite{bappy2019hybrid} & \underline{.255} & .068 & .181 & .042 & .822 & .713 & .262  & .069 \\
\midrule  
 SPAN \cite{hu2020span} & .131 & .178 & .070 & .098 & .905 & .859 & .122 & .178 \\
\midrule  
  ViT-B \cite{dosovitskiy2020vit}  & .245 & .156 & .142 & .085 & .862 & .804 & .241  & .161 \\
\midrule  
  Swin-ViT \cite{liu2021swin}  & .214 & .174 & .120 & .095 & \underline{.923} & .903 &  .232 & .170 \\
\midrule 
 PSCC \cite{liu2022pscc} & .020 & .013 & .010 & .007 & .609 & .547 & .000 & .000\\
\midrule  
 MVSS-Net++ \cite{dong2022mvss}  & .067 & \underline{.264} & .035 & \underline{.152} & .741 & 889 & .063 & \underline{.261} \\
\midrule  
 CAT-NET \cite{kwon2022learning}  & .089 & .178 & .068 & .141 & .588 & .787 & .088 & .185 \\
\midrule 
 EVP \cite{liu2023evp}  & .028 & .164 & .014 & .089 & .916 & \textbf{.923} & .074 & .196 \\
\midrule  
 TruFor \cite{guillaro2023trufor} & .234 & .221 & .133 & .124 & .891 & .875 & .249 & .240 \\
\midrule[1.5pt]
 \cellcolor[HTML]{E0DBDB}PIM & \cellcolor[HTML]{E0DBDB}\textbf{.357} & \cellcolor[HTML]{E0DBDB}\textbf{.288} & \cellcolor[HTML]{E0DBDB}\textbf{.217} & \cellcolor[HTML]{E0DBDB}\textbf{.168} & \cellcolor[HTML]{E0DBDB}\textbf{.953} & \cellcolor[HTML]{E0DBDB}\underline{.914} & \cellcolor[HTML]{E0DBDB}\textbf{.351} & \cellcolor[HTML]{E0DBDB}\textbf{.300} \\
\bottomrule
\end{tabular}} 										
\end{table}

\begin{table*}
  \centering
  \caption{\textcolor{black}{Image manipulation localization performance on unseen advanced manipulation techniques.}}
  \scalebox{0.76}{\begin{tabular}{lccccccccccc}
    \toprule
    \multirow{2}{*}{Method} & \multirow{2}{*}{Venue} & \multicolumn{2}{c}{AutoSplice 100} & \multicolumn{2}{c}{AutoSplice 90} & \multicolumn{2}{c}{AutoSplice 75} & \multicolumn{2}{c}{CocoGlide} & \multicolumn{2}{c}{Avgerage} \\ \cline{3-12}
    & & AUC & MCC & AUC & MCC & AUC & MCC & AUC & MCC & AUC & MCC\\ \midrule
    FCN \cite{long2015fully} & CVPR15 & .681 & .150 & .619 & .078 & .589 & .048 & .618 & .079 & .627 & .089 \\
    \midrule 							
    U-Net \cite{ronneberger2015u} & MICCAI15 & .616 & .072 & .585 & .047 & .570 & .034 & .578 & .047 & .587 & .050 \\
    \midrule 						
    DeepLabv3 \cite{chen2017deeplab} & TPAMI18 & .864 & .223 & .812 & .153 & \underline{.759} & .099 & .730 & .103 & .791 & .145 \\
    \midrule	 					
    MFCN \cite{salloum2018image} & JVCIP18 & .565 & .072 & .547 & .049 & .534 & .035 & .551 & .052 & .549 & .052 \\
    \midrule 	 							
    RRU-Net \cite{bi2019rru} & CVPRW19 & .781 & .159  & .737 & .114 & .714 & .083 & .620 & .051 & .713 & .102 \\  
    \midrule	 								
    MantraNet \cite{wu2019mantra} & CVPR19 & .664 & .189 & .626 & .160 & .660 & \underline{.177} & .806 & .190 & .689 & .179 \\
    \midrule	 				
    HPFCN \cite{li2019localization} & ICCV19 & .646 & .092 & .633 & .082 & .622 & .067 & .586 & .048 & .622 & .072 \\  
    \midrule 								
    H-LSTM \cite{bappy2019hybrid} & TIP19 & .639 & .162 & .643 & .162 & .634 & .145 & .643 & .137  & .640 & .152 \\
    \midrule  
    SPAN \cite{hu2020span} & ECCV20 & .645 & .020 & .549 & .005 & .566 & .007 & .776	& .198 & .634 & .058 \\
    \midrule						
    ViT-B \cite{dosovitskiy2020vit} & ICLR21 & .662 & .131 & .658 & .126 & .651 & .118 & .631 & .105 & .651 & .120 \\
    \midrule 						
    Swin-ViT \cite{liu2021swin} & ICCV21 & .700 & .233 & .590 & .072 & .570 & .046 & .648 & .126 & .627 & .119 \\
    \midrule 	 				
    PSCC \cite{liu2022pscc} & TCSVT22 & .749 & .275 & .657 & .195 & .630 & .156 & .566 & .051 & .651 & .169 \\  
    \midrule	  						
    MVSS-Net++ \cite{dong2022mvss} & TPAMI22 & .836 & .280 & .751 & .101 & .714 & .054 & \textbf{.819} & \underline{.309} & .780 & .186 \\
    \midrule	 					
    CAT-NET \cite{kwon2022learning} & IJCV22 & \underline{.887} & \underline{.578} & .720 & .301 & .607 & .163 & .587 & .133 & .700 & .294 \\
    \midrule	 
    EVP \cite{liu2023evp} & CVPR23 & .762 & .226 & .697 & .124 & .637 & .078 & .686 & .114 & .696 & .136 \\ 
    \midrule							
    TruFor \cite{guillaro2023trufor} & CVPR23 & .827 & .382 & \underline{.818} & \underline{.358} & \textbf{.820} & \textbf{.367} & .757 & .253 & \underline{.806} & \underline{.340} \\
    \midrule[1.5pt]	
    \cellcolor[HTML]{E0DBDB}PIM & \cellcolor[HTML]{E0DBDB}Ours & \cellcolor[HTML]{E0DBDB}\textbf{.940} & \cellcolor[HTML]{E0DBDB}\textbf{.715} & \cellcolor[HTML]{E0DBDB}\textbf{.852} & \cellcolor[HTML]{E0DBDB}\textbf{.402} & \cellcolor[HTML]{E0DBDB}{.729} & \cellcolor[HTML]{E0DBDB}{.151} & \cellcolor[HTML]{E0DBDB}\underline{.817} & \cellcolor[HTML]{E0DBDB}\textbf{.372} & \cellcolor[HTML]{E0DBDB}\textbf{.835} & \cellcolor[HTML]{E0DBDB}\textbf{.410}\\
    \bottomrule					
  \end{tabular}}
  \label{advanced}
\end{table*}

\begin{figure*}[ht]
\centering
\includegraphics[scale=0.393]{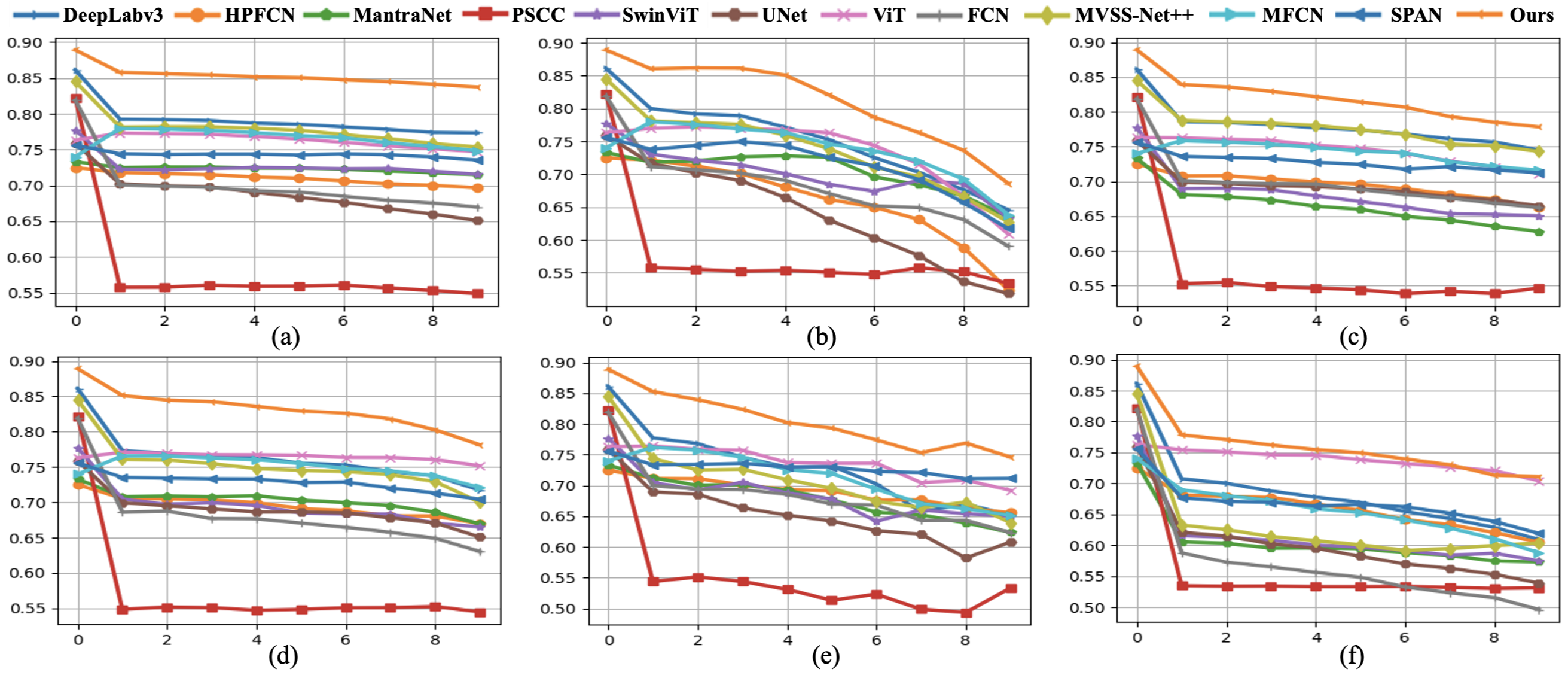}
\caption{Robustness evaluation results (AUC) on six unseen perturbation types: (a). Brightness, (b). Contrast, (c). Darkening, (d). Dithering, (e). Pink noise, (f). JPEG2000. The x-axis indicates the perturbation severity level. 
}
\label{Robustness_AUC}
\end{figure*}

\subsection{Robustness Evaluation Results} 
Due to uncontrollable variables in real-world applications (e.g., black-box compression via social media platforms), detectors may encounter unseen image perturbations, resulting in significant performance drops. Although regular data augmentations have been considered during the training process, it is challenging to foresee all perturbation types under the deployment circumstance.

This study introduced six common image perturbations, brightness, contrast, darkening, dithering, pink noise, and JPEG2000 compression, on the CASIAv1+ \cite{dong2013casia} dataset, which was unknown during the training process. We further set nine severity levels for each perturbation type to accommodate various environmental variations. \textcolor{black}{We showcase examples of raw images and the corresponding perturbed versions in the Appendix.} The pixel-level AUC detection scores are illustrated in Fig.~\ref{Robustness_AUC}. The $x$ dimension indicates the severity levels, where Severity `0' indicates no perturbation applied. We can observe that all detection models suffer certain performance drops due to these unforeseen perturbation types. The proposed method consistently achieves the best AUC across different perturbation levels on all unseen perturbation types, demonstrating the robustness of our method. As most image perturbations encountered in real-world scenarios are uniformly applied to images, the pixel dependencies within unaltered images and the pixel inconsistencies within manipulated images remain consistent. Therefore, our proposed method continues to exhibit the best forgery localization performance in such robustness evaluations.



\begin{table*}
  \caption{Ablation study for image manipulation localization. }
  \label{ablation}
  \centering
  \renewcommand\arraystretch{1.15}
  \scalebox{0.92}{\begin{tabular}{ccccccccccccc}
    \hline
    Setting & BB & BDD & RDA & PIDA & CDC & RDC & LWM & $L_{C}$ & GPDE & $L_{R}$ & AVG. F1 & AVG. IoU \\
    \hline
    \hline
    1 & \checkmark & - & - & - & - & - & - & - & - & - & .211 & .171 \\
    2 & \checkmark & \checkmark & - & - & - & - & - & - & - & - & .220 & .178 \\
    3 & \checkmark & \checkmark & \checkmark & - & - & - & - & - & - & - & .233 & .190 \\
    4 & \checkmark & \checkmark & - & \checkmark & - & - & - & - & - & - & .260 & .209 \\
    5 & \checkmark & \checkmark & \checkmark & \checkmark & - & - & - & - & - & - & .283 & .237 \\
    6 & \checkmark & \checkmark & \checkmark & \checkmark & \checkmark & - & - & - & - & - & .304 &  .252 \\
    7 & \checkmark & \checkmark & \checkmark & \checkmark & - & \checkmark & - & - & - & - & .308 & .258 \\
    8 & \checkmark & \checkmark & \checkmark & \checkmark & \checkmark & \checkmark & - & - & - & - & .312 & .262 \\
    9 & \checkmark & \checkmark & \checkmark & \checkmark & \checkmark & \checkmark & \checkmark & - & - & - & .317 & {.271}  \\
    10 & \checkmark & \checkmark & \checkmark & \checkmark & \checkmark & \checkmark & \checkmark & \checkmark & - & - & {.323} & .269  \\
    11 & \checkmark & \checkmark & \checkmark & \checkmark & \checkmark & \checkmark & \checkmark & \checkmark & \checkmark & - & .330 & .272 \\
    \hline
    12 & \checkmark & \checkmark & \checkmark & \checkmark & \checkmark & \checkmark & \checkmark & \checkmark & \checkmark & \checkmark & \textbf{.333} & \textbf{.275} \\
    \hline
\end{tabular}}
\end{table*}

\begin{figure*}[ht]
\centering
\includegraphics[scale=0.46]{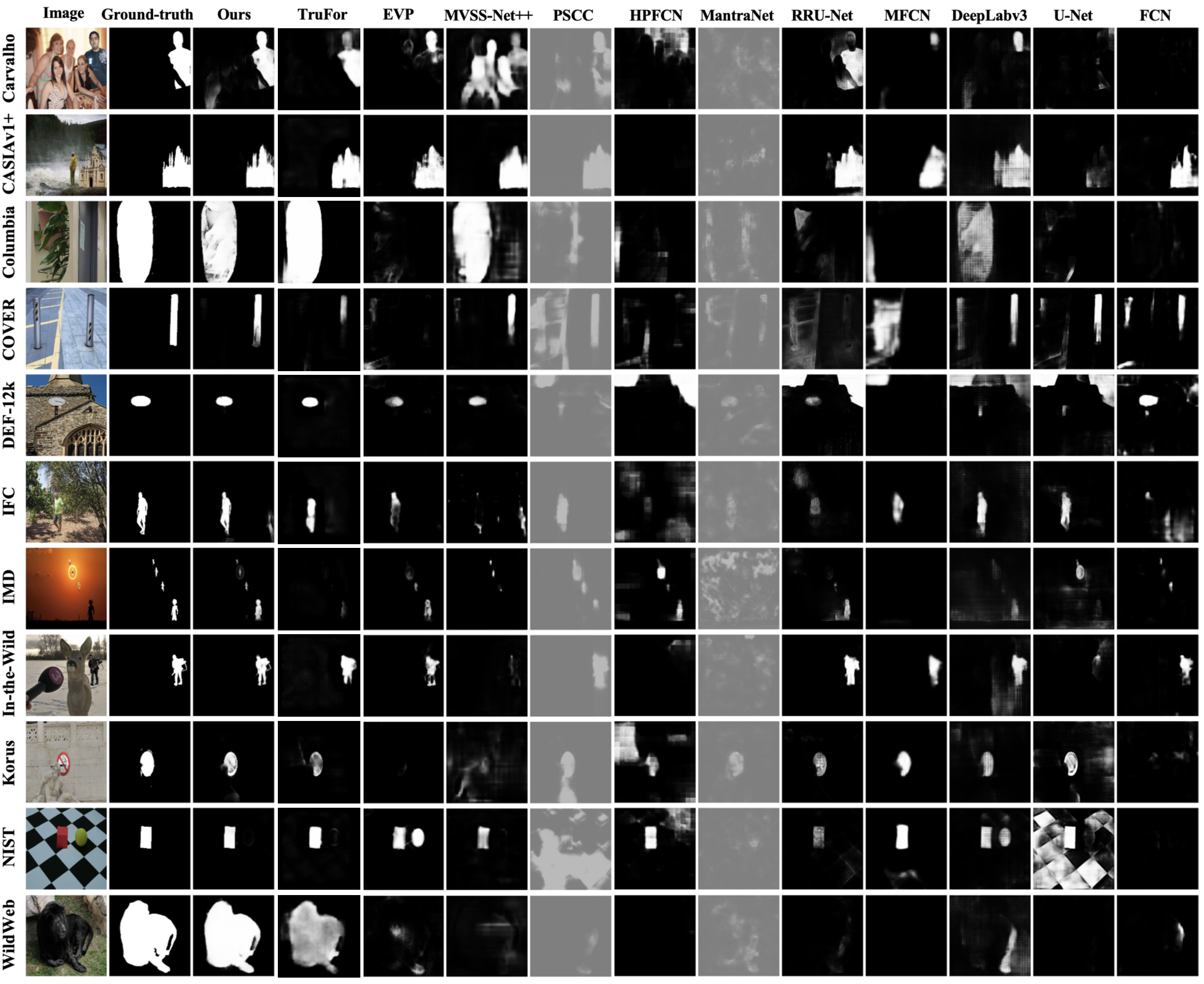}
\caption{\textcolor{black}{Forgery localization results on the 11 unseen test sets. The three left columns show the input images, corresponding ground-truth, and the localization results of our method. The right 11 columns present the results of SOTA methods.}
}
\label{visualization}
\end{figure*}

\begin{figure*}[ht]
\centering
\includegraphics[scale=0.34]{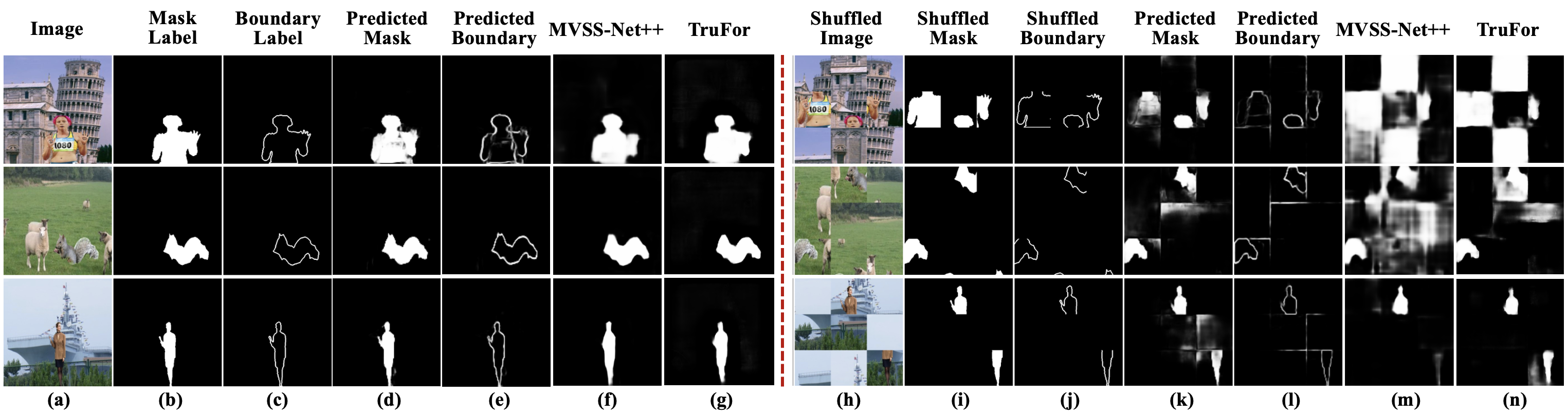}
\caption{Visualization results on shuffled images. (a) Input unshuffled images. (b) Forgery localization labels. (c) Forgery boundary labels. (d) Our forgery localization results. (e) Our boundary prediction results. (f) MVSS-Net++ forgery localization results. 
\textcolor{black}{(g) TruFor forgery localization results.} (h) Input shuffled images. (i) Shuffled forgery localization labels. (j) Shuffled forgery boundary labels.  (k) Our forgery localization results on shuffled images. (l) Our boundary prediction results on shuffled images. (m) MVSS-Net++ forgery localization results on shuffled images. \textcolor{black}{(n) TruFor forgery localization results on shuffled images.} 
}
\label{shuffle}
\end{figure*}

\subsection{Qualitative Experimental Results} 
In Fig.~\ref{visualization}, we qualitatively evaluate the image manipulation localization performance across 11 unseen test sets, where the leftmost three columns show the input images, the corresponding ground-truth masks, and the predicted results of our method. Besides, we show the forgery localization results of SOTA methods in the right 11 columns. Our method can accurately localize the manipulated regions for forgery images with diverse image quality, scenes, occlusions, and illumination conditions. Our localization results are superior to previous methods, regardless of whether the forgery regions are relatively substantial (e.g., Columbia and WildWeb) or subtle (e.g., DEF-12k and Korus) in fake images.

The boundaries of predicted results are much sharper for the proposed method than in previous arts. This can be attributed to the global pixel dependency modeling module and local pixel difference convolution module that can highlight pixel inconsistency in forgery boundary regions. As the predicted results in CASIAv1+ and In-the-Wild show, the proposed method can successfully localize extremely subtle forgery details. This can be attributed to the local pixel difference convolution module, which allows the model to capture local pixel inconsistency artifacts. Our method maintains accurate localization performance for more challenging images, such as the one in the IMD row that contains multiple tiny forgery regions. Finally, the proposed method results in fewer false alarms, as evidenced in the predictions of COVER and NIST. This characteristic can ensure a more dependable forgery detection for real-world deployment. \textcolor{black}{Compared to TruFor, our method PIM exhibits more accurate forgery localization, fewer false alarms, sharper forgery boundaries, and superior capability in capturing subtle forgery traces.}
We provide more visualization results in the Appendix.

The qualitative experimental results demonstrate that the proposed formulation effectively deals with various challenging forgery situations. This is primarily attributed to the dedicated module designs to extract inherent pixel-level forgery fingerprints. 
\subsection{Visualization Results on Shuffled Images}
To demonstrate the effectiveness of the proposed model in capturing pixel inconsistency artifacts for forgery localization, we split the input image into 3$\times$3 patches and shuffle them randomly. This random shuffling effectively suppresses the semantic information within the input images and allows us to assess whether our model can still accurately localize the forgery regions. We present results for unshuffled and shuffled images in Fig.~\ref{shuffle}, denoted as (a)-(g) and (h)-(n), respectively. Columns (a)-(c) show the original input images, their respective mask, and boundary labels. 
\textcolor{black}{Columns (d)-(g) present our forgery localization maps, boundary predictions, localization results of MVSS-Net++, and localization results of TruFor. In this evaluation, we select the MVSS-Net++ and TruFor as the baselines as they are the SOTA forgery localization methods according to our experimental results in Sec. 4.5-4.8. We observe that PIM (Ours), MVSS-Net++, and TruFor can successfully predict manipulated regions in the unshuffled images.}  

Next, we present  prediction results on shuffled images in Fig.~\ref{shuffle} (h)-(n). These randomly shuffled images inherently contain limited semantic information. In column (k), the proposed method effectively localizes the forgery regions within each patch. Column (l) showcases the predicted sharp boundaries of forgery patches. In contrast, forgery prediction results of MVSS-Net++ in column (m) reveal struggling performance, marked by numerous false alarms and undetected forgery regions. \textcolor{black}{While TruFor aims to capture generic noise artifacts in forgery images, column (n) shows that it still performs poorly in such a challenging setting.} The localization results of shuffled images further demonstrate the superiority of our method. Therefore, we conclude the proposed method focuses more on pixel-level artifacts than semantic-level forgery traces.



\subsection{Ablation Experiments} 
In this subsection, we present comprehensive ablation studies to evaluate the effectiveness of the components designed in our framework. Table.~\ref{ablation} shows the average forgery localization performance in the cross-dataset evaluations, where `$\checkmark$' denotes the used component. 

BB indicates the ensemble of the transformer backbone and the mask decoder. BDD denotes the utilization of the boundary decoder. RDA and PIDA represent the regular data augmentation and the proposed Pixel-Inconsistency Data Augmentation. CDC, RDC, and LWM stand for central pixel difference convolution, radial pixel difference convolution, and the learning to weight module, respectively. $L_{C}$ indicates the usage of the compactness loss. \textcolor{black}{GPDE and $L_{R}$ represent the designed Global Pixel Dependency Encoder and the reconstruction loss, respectively.}

From Table.~\ref{ablation}, we can observe that using a boundary decoder can boost the forgery localization performance. A comparison between Setting 3 and 4 highlights the superiority of the proposed PIDA over RDA, suggesting that PIDA encourages the detector to focus on more general artifacts. Intuitively, the combination of RDA and PIDA in Setting 5 is expected to enhance pixel-level forgery detection performance, primarily because the model has been fed more data.
The CDC and RDC modules (Settings 6-8) effectively capture local pixel difference features, contributing to enhanced localization results. Furthermore, Setting 9 demonstrates the effectiveness of the LWM, which learns the weights more smartly and performs a better feature fusion. Using the compactness loss $L_C$ in Setting 10 produces more compact outputs, improving the final performance. 
\textcolor{black}{The use of GPDE in Setting 11 successfully models global pixel dependency, thereby achieving superior image forgery localization performance. Compared to Setting 11, Setting 12 adopts the reconstruction loss $L_{R}$ to further enhance global pixel dependency modeling while revealing pixel inconsistency artifacts in manipulated images. This contributes significantly to the overall localization performance. The detailed ablation experimental results across all testing datasets, the experiments regarding the impacts of multi-head self-attention, and the visualization ablation results can be found in the Appendix.} 

In summary, the ablation studies exhibit the critical role of the designed components in our framework. The ensemble of these components jointly enhances the forgery localization performance.

\vspace{-3mm}
\section{Conclusions and Future Work}
This paper presented a generalized and robust image manipulation localization model by capturing pixel inconsistency in forgery images. The method is underpinned by a two-stream pixel dependency modeling framework for image forgery localization. It incorporates a novel masked self-attention mechanism to model the global pixel dependencies within input images effectively. Additionally, two customized convolutional modules, the Central Difference Convolution (CDC) and the Radial Difference Convolution (RDC), better capture pixel inconsistency artifacts within local regions. We find that modeling pixel interrelations can effectively mine intrinsic forgery clues. To enhance the overall performance, Learning-to-Weight Modules (LWM) complementarily combines global and local features. The usage of the dynamic weighting scheme can lead to a better feature fusion, contributing to a more robust and generalized image forgery localization.

Furthermore, a novel Pixel-Inconsistency Data Augmentation (PIDA) that exclusively employs pristine images to generate augmented forgery samples, guides the focus on pixel-level artifacts. The proposed PIDA strategy can shed light on improving the generalization for future forensics research. Extensive experimental results demonstrated the state-of-the-art performance of the proposed framework in image manipulation detection and localization, both in generalization and robustness evaluations. \textcolor{black}{Our designed model also exhibits outstanding performance on unseen, advanced, and sophisticated manipulation images, underscoring its potential in challenging real-world scenarios.} The ablation studies further validated the effectiveness of the designed components. 


While our method is robust against unseen image perturbations, it remains susceptible to recapturing attacks. This vulnerability stems from the framework's primary objective: to identify pixel inconsistency artifacts resulting from the disruption of CFA regularity during the manipulation process. Recapturing operations reintroduce the pixel dependencies initially constructed during the demosaicing process, concealing the pixel inconsistency artifacts and leading to failed forgery detection. In future research, developing an effective recapturing detection module becomes a crucial research direction to ensure more secure manipulation detection.




\vspace{-6mm}
{\small
\bibliographystyle{plain}
\bibliography{main}
}

\begin{IEEEbiography}[{\includegraphics[width=1in,height=1.25in,clip,keepaspectratio]{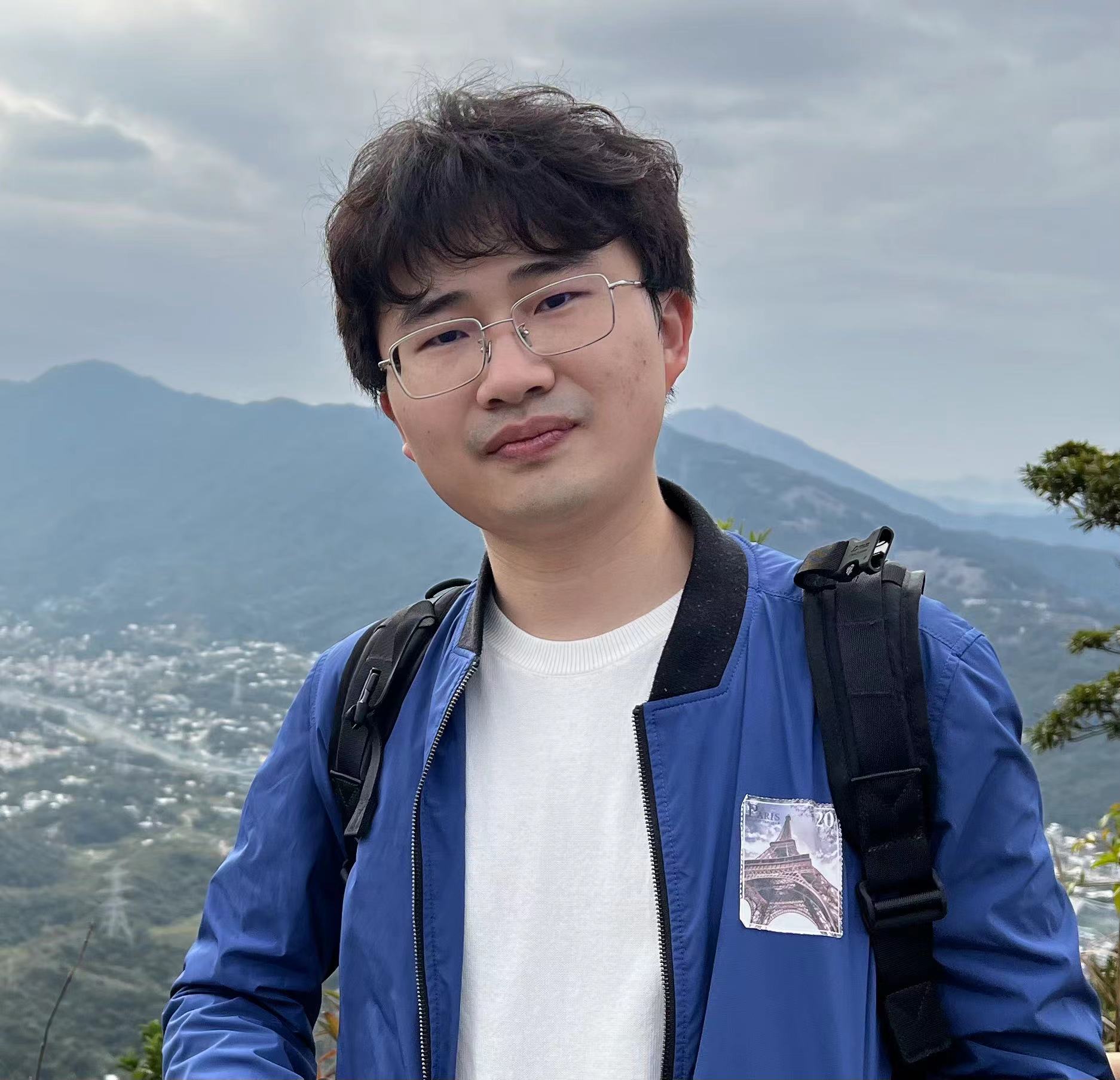}}]{Chenqi Kong} received the B.S. and M.S. degrees in the College of Science and the College of Electrical Engineering and Automation, Harbin Institute of Technology, Harbin, China, in 2017 and 2019, respectively. He received the Ph.D. degree in the Department of Computer Science, City University of Hong Kong, Hong Kong, China (Hong Kong SAR) in 2023. He is currently a research fellow in the School of Electrical and Electronic Engineering, Nanyang Technological University, Singapore. He is a recipient of National Scholarship and  Research Tuition Scholarship. His research interests include AI security and multimedia forensics.
\end{IEEEbiography}

\begin{IEEEbiography}[{\includegraphics[width=1in,height=1.25in,clip,keepaspectratio]{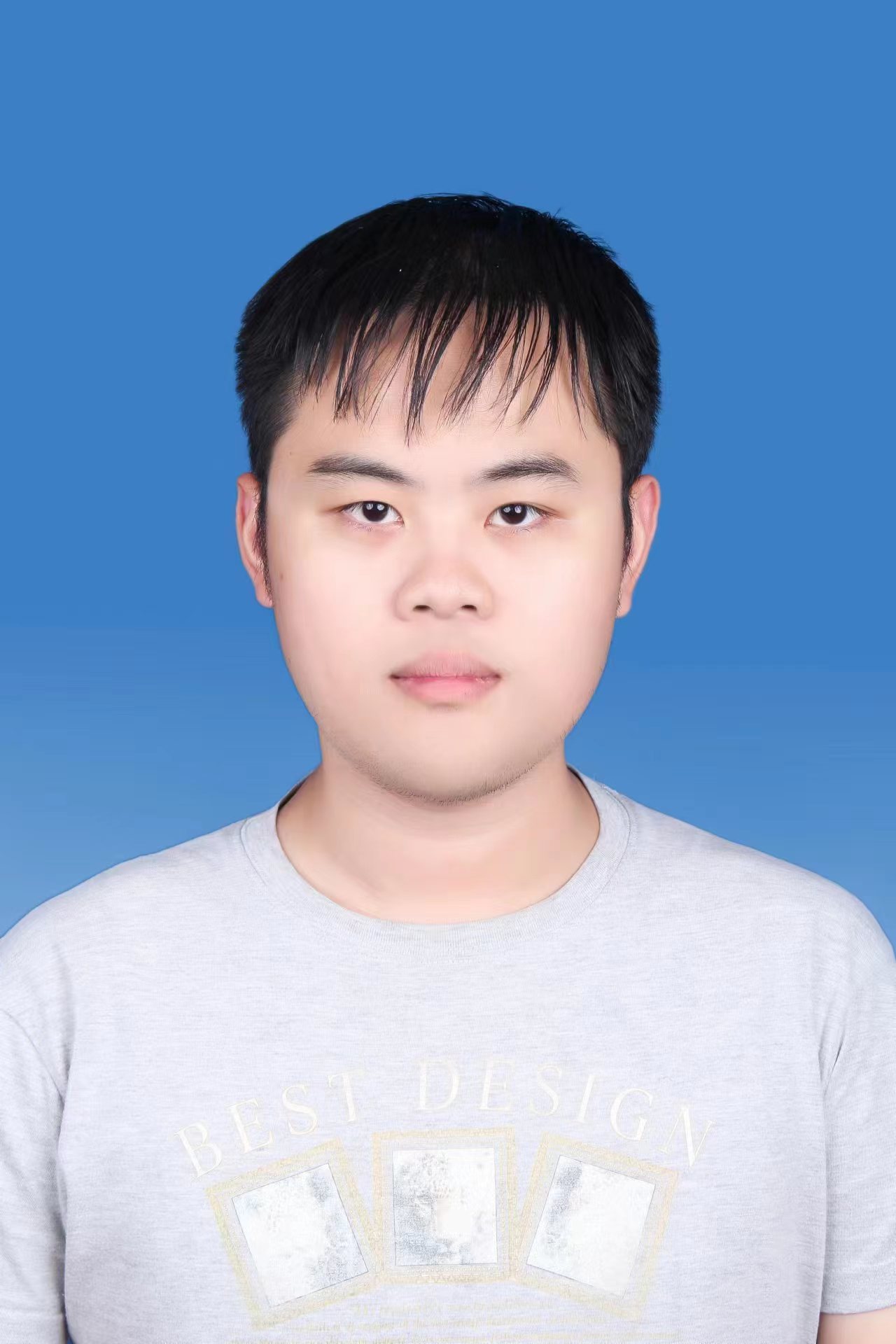}}]{Anwei Luo} received the B.S. degree from Jilin University, Changchun, China, in 2013. He is currently pursuing the Ph. D. degree from Sun Yat-sen University, Guangzhou, China. His current research interests include digital multimedia forensics, watermarking and security.
\end{IEEEbiography}

\begin{IEEEbiography}[{\includegraphics[width=1in,height=1.25in,clip,keepaspectratio]{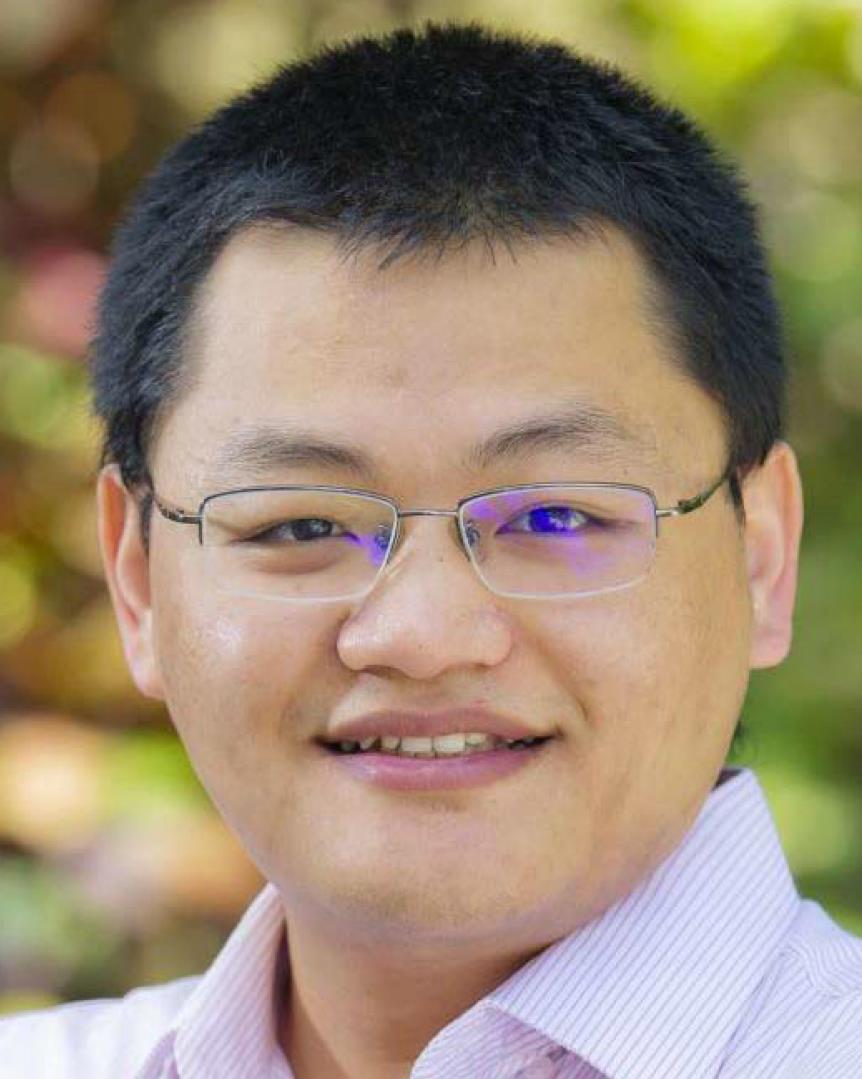}}]{Shiqi Wang} received the B.S. degree in computer science from the Harbin Institute of Technology in 2008 and the Ph.D. degree in computer application technology from Peking University in 2014. From 2014 to 2016, he was a Post-Doctoral Fellow with the Department of Electrical and Computer Engineering, University of
Waterloo, Waterloo, ON, Canada. From 2016 to
2017, he was a Research Fellow with the Rapid-Rich
Object Search Laboratory, Nanyang Technological
University, Singapore. He is currently an Associate Professor with the Department of Computer Science, City University of Hong Kong. He has proposed over 40 technical proposals to ISO/MPEG, ITU-T, and AVS standards, and authored/coauthored more than 200 refereed journal articles/conference papers. He received the Best Paper Award from IEEE VCIP 2019, ICME 2019, IEEE Multimedia 2018, and PCM 2017 and is the coauthor of an article that received the Best Student Paper Award in the IEEE ICIP 2018. His research interests include video compression, image/video quality assessment, and image/video search and analysis. 
\end{IEEEbiography}

\vspace{-1cm}
\begin{IEEEbiography}[{\includegraphics[width=1in,height=1.25in,clip,keepaspectratio]{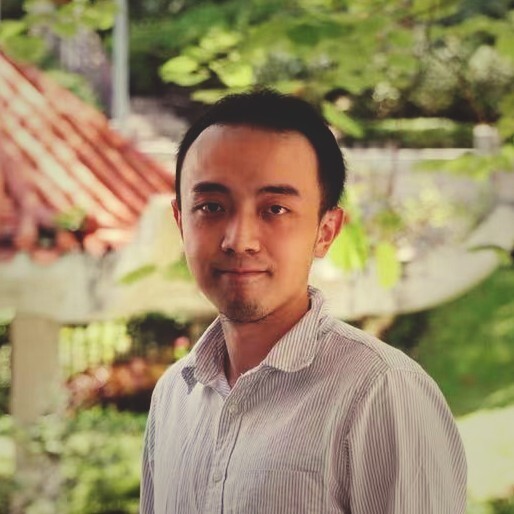}}]{Haoliang Li} received the B.S. degree in communication engineering from University of Electronic Science and Technology of China (UESTC) in 2013, and his Ph.D. degree from Nanyang Technological University (NTU), Singapore in 2018. He is currently an assistant professor in Department of Electrical Engineering, City University of Hong Kong. His research mainly focuses on AI security, multimedia forensics and transfer learning. His research works appear in international journals/conferences such as TPAMI, IJCV, TIFS, NeurIPS, CVPR and AAAI. He received the Wallenberg-NTU presidential postdoc fellowship in 2019, doctoral innovation award in 2019, and VCIP best paper award in 2020.  
\end{IEEEbiography}

\begin{IEEEbiography}[{\includegraphics[width=1in,height=1.25in,clip,keepaspectratio]{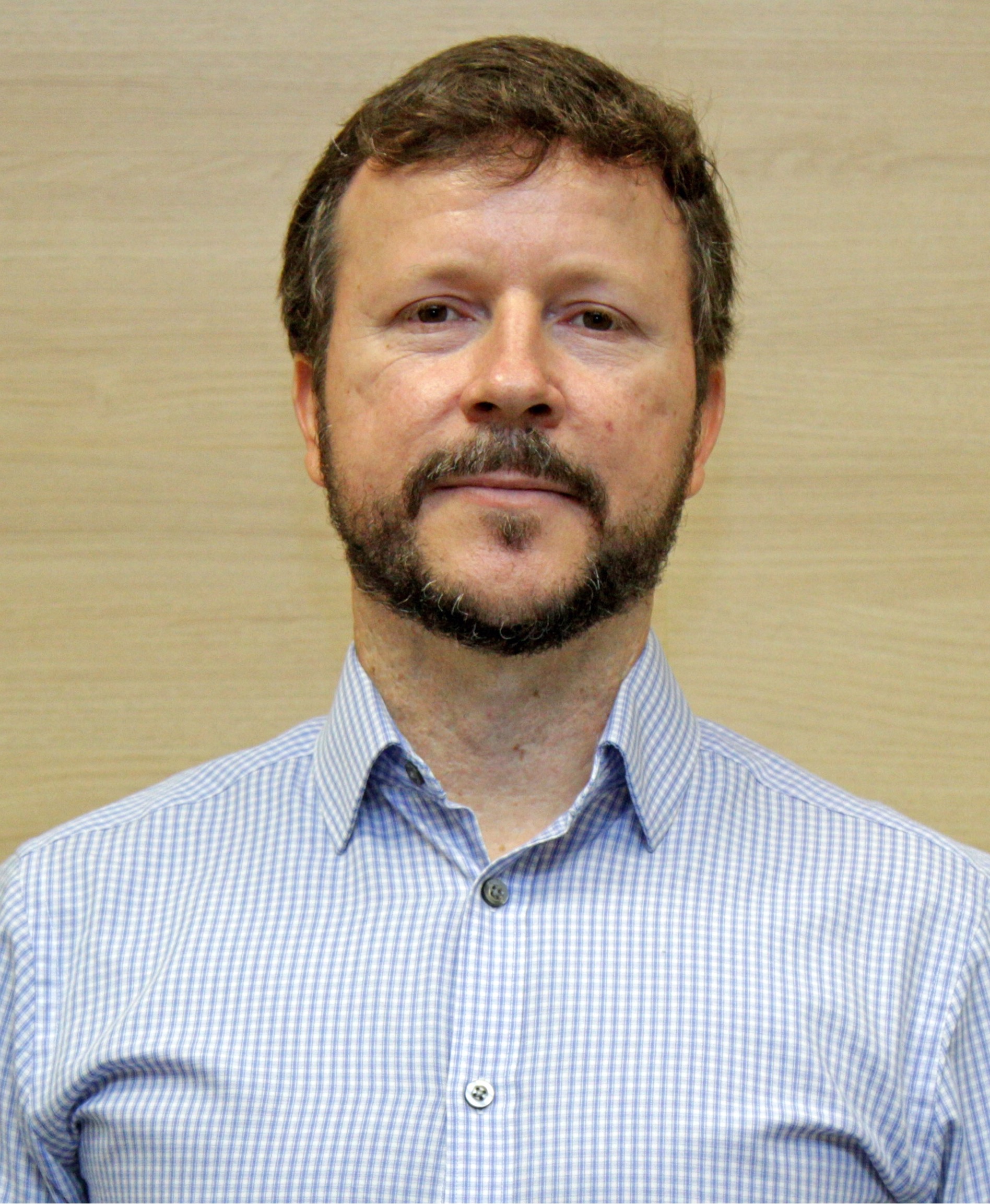}}]{Anderson Rocha} received his Ph.D. degree in computer science. He is a full professor of artificial intelligence and digital forensics at the Institute of Computing, University of Campinas, Campinas 13083-852, Brazil, where he is the coordinator of the Artificial Intelligence Lab. A Microsoft and Google Faculty Fellow, he is a former chair of the IEEE Information Forensics and Security Technical Committee (2019–2020) and an affiliated member of the Brazilian Academy of Sciences and the Brazilian Academy of Forensics Sciences. His research interests include artificial intelligence, digital forensics, and reasoning for complex data. He is a Fellow of IEEE.
\end{IEEEbiography}

\begin{IEEEbiography}[{\includegraphics[width=1in,height=1.25in,clip,keepaspectratio]{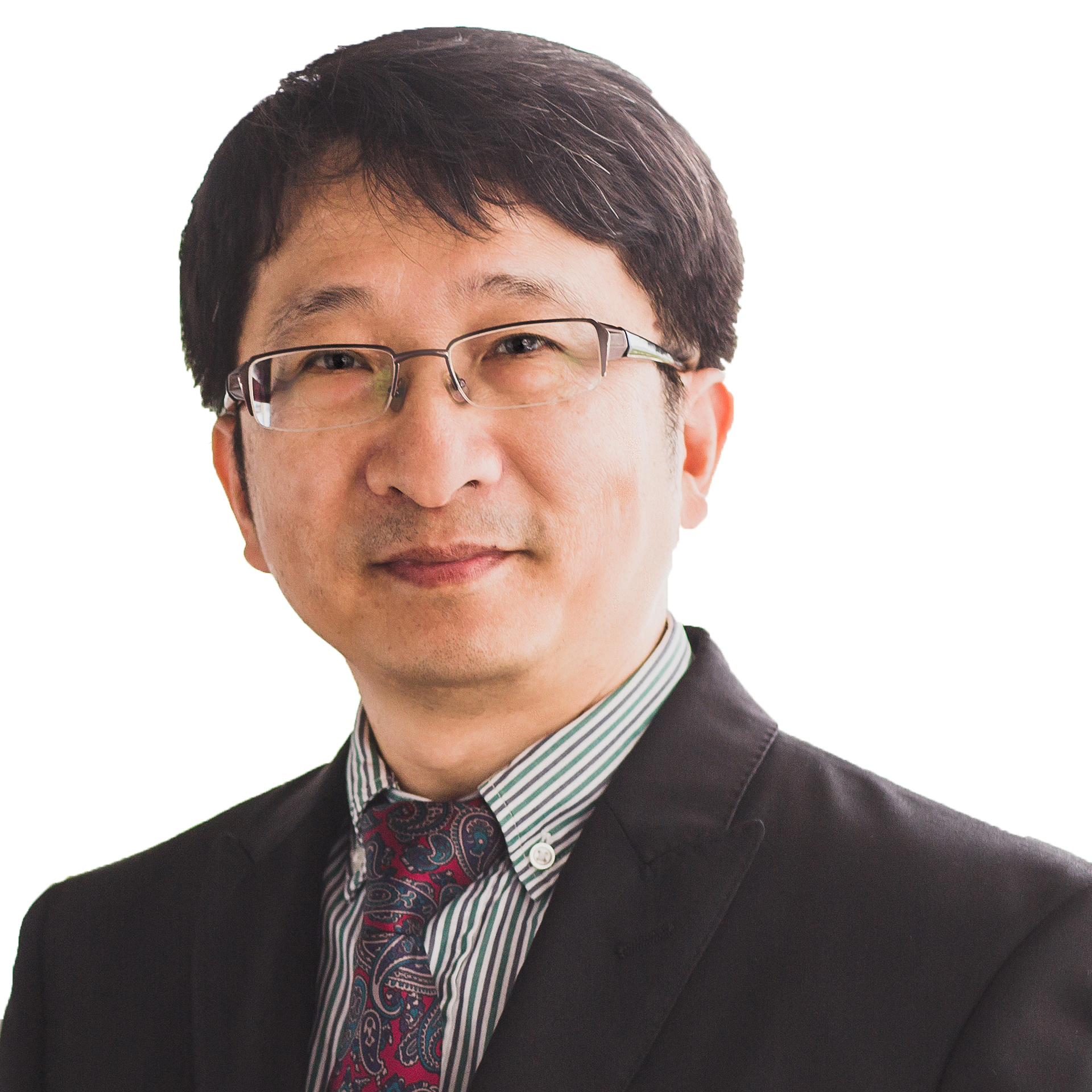}}]
{Prof. Alex Kot} has been with the Nanyang Technological University, Singapore since 1991. He was Head of the Division of Information Engineering and Vice Dean Research at the School of Electrical and Electronic Engineering. Subsequently, he served as Associate Dean for College of Engineering for eight years. He is currently Professor and Director of Rapid-Rich Object SEarch (ROSE) Lab and NTU-PKU Joint Research Institute. He has published extensively in the areas of signal processing, biometrics, image forensics and security, and computer vision and machine learning.
Prof. Kot served as Associate Editor for more than ten journals, mostly for IEEE transactions. He served the IEEE SP Society in various capacities such as the General Co-Chair for the 2004 IEEE International Conference on Image Processing and the Vice-President for the IEEE Signal Processing Society. He received the Best Teacher of the Year Award and is a co-author for several Best Paper Awards including ICPR, IEEE WIFS and IWDW, CVPR Precognition Workshop and VCIP. He was elected as the IEEE Distinguished Lecturer for the Signal Processing Society and the Circuits and Systems Society. He is a Fellow of IEEE, and a Fellow of Academy of Engineering, Singapore. 
\end{IEEEbiography}

\begin{table*}
  \centering
  \vspace{-1.5mm}
  \caption{\textcolor{black}{Image manipulation localization performance (\textbf{AUC score}).}}
  \vspace{-2mm}
  \scalebox{0.67}{\begin{tabular}{lcccccccccccccc}
    \toprule
    Method & Venue & NIST & Columbia & CASIAv1+ & COVER & DEF-12k & IMD & Carvalho & IFC & In-the-Wild & Korus & WildWeb & AVG \\
    \midrule
    FCN \cite{long2015fully} & CVPR15 & .675 & .696 & .819 & .694 & .628 & .748 & .686 & .605 & .690 & .644 & .651 & .685\\ 
    \midrule
    U-Net \cite{ronneberger2015u} & MICCAI15 & .668 & .645 & .759 & .622 & .587 & .703 & .653 & .598 & .654 & .626 & .591 & .646\\ 										
    \midrule 	
    DeepLabv3 \cite{chen2017deeplab} & TPAMI18 & .720 & .853 & \underline{.861} & .763 & .667 & .815 & \textbf{.807} & .631 & {.752} & .675 & {.709} & .750\\ 											
    \midrule
    MFCN \cite{salloum2018image} & JVCIP18 & .691 & .634 & .740 & .614 & .576 & .664 & .631 & .591 & .621 & .621 & .575 & .633 \\ 
    \midrule 	
    RRU-Net \cite{bi2019rru} & CVPRW19 & .715 & .749 & .800 & .676 & .593 & .754 & .661 & .586 & .704 & .669 & .633 & .685  \\ 
    \midrule 											
    MantraNet \cite{wu2019mantra} & CVPR19 & .734 & .734 & .733 & .722 & .696 & .760 & .644 & .592 &  .719 & .646 & .626 & .691\\			
    \midrule
    HPFCN \cite{li2019localization} & ICCV19 & .688 & .607 & .725 & .591 & .583 & .683 & .583 & .564 & .642 & .607 & .626 & .627\\		
    \midrule 
    H-LSTM \cite{bappy2019hybrid} & TIP19 & .696 & .571 & .634 & .634 & .581 & .656 & .586 & .553 & .611 & .588 & .630 & .613\\ 
    \midrule 
    SPAN \cite{hu2020span} & ECCV20 & .751 & {.855} & .756 & .777 & .641 & .763 & .671 & .602 & .749 & .649 & .582 & .709\\ 
    \midrule							
    ViT-B \cite{dosovitskiy2020vit} & ICLR21 & .705 & .689 & .763 & .665 & .602 & .693 & .674 & .580 & .692 & .653 & .605 & .666 \\   
    \midrule 											
    Swin-ViT \cite{liu2021swin} & ICCV21 & .723 & .750 & .777 & .740 & .669 & .793 & .668 & .641 & .710 & {.701} & .572 & .704 \\  								
    \midrule
    PSCC \cite{liu2022pscc} & TCSVT22 & .676 & .731 & .822 & .660 & .600 & .762 & .700 & .589 & .696 & .646 & .558 & .676\\ 								
    \midrule									
    MVSS-Net++ \cite{dong2022mvss} & TPAMI22 & \textbf{.791} & .818 & .845 & \textbf{.871} & .683 & {.817} & .731 & .635 & {.794} & .659 & .646 & {.754}\\	
    \midrule	
    CAT-NET \cite{kwon2022learning} & IJCV22 & .522 & .524 & .668 & .662 & \textbf{.818} & .588 & .603 & .442 & .504 & .531 & .536 & .582\\	
    \midrule 									
    EVP \cite{liu2023evp} & CVPR23 & \underline{.775} & .791 & .855 & .716 & \underline{.697} & .811 & .688 & \underline{.648} & .748 & \underline{.715} & .695 & .740 \\
     \midrule 
    \textcolor{black}{TruFor} \cite{guillaro2023trufor} & CVPR23 & .745 & \textbf{.916} & \textbf{.889} & \underline{.827} & .629 & \underline{.832} & .739 & .634 & \underline{.802} & .670 & \underline{.724} & \underline{.764} \\
    \midrule[1.5pt]
     \cellcolor[HTML]{E0DBDB}PIM & \cellcolor[HTML]{E0DBDB}Ours & \cellcolor[HTML]{E0DBDB}{.752} & \cellcolor[HTML]{E0DBDB}\underline{.884} & \cellcolor[HTML]{E0DBDB}\textbf{.889} & \cellcolor[HTML]{E0DBDB}.809 & \cellcolor[HTML]{E0DBDB}.687 & \cellcolor[HTML]{E0DBDB}\textbf{.870} & \cellcolor[HTML]{E0DBDB}\underline{.760} & \cellcolor[HTML]{E0DBDB}\textbf{.669} & \cellcolor[HTML]{E0DBDB}\textbf{.831} & \cellcolor[HTML]{E0DBDB}\textbf{.725} & \cellcolor[HTML]{E0DBDB}\textbf{.725} & \cellcolor[HTML]{E0DBDB}\textbf{.782} \\
    \bottomrule
  \end{tabular}}
  \label{auc_table}
\end{table*}

\vspace{-2mm}
\begin{table*}
  \centering
  \caption{\textcolor{black}{Image manipulation localization performance (\textbf{MCC score} with fixed threshold: 0.5).}}
  \vspace{-2mm}
  \scalebox{0.67}{\begin{tabular}{lcccccccccccccc}
    \toprule
    Method & Venue & NIST & Columbia & CASIAv1+ & COVER & DEF-12k & IMD & Carvalho & IFC & In-the-Wild & Korus & WildWeb & AVG \\
    \midrule
    FCN \cite{long2015fully} & CVPR15 & .151 & .194 & .425 & .154 & .113 & .212 & .083 & .078 & .192 & .126 & .162 & .172 \\
    \midrule 
    U-Net \cite{ronneberger2015u} & MICCAI15 & .155 & .119 & .263 & .073 & .036 & .137 & .098 & .058 & .140 & .105 & .053 & .112 \\
    \midrule 	
    DeepLabv3 \cite{chen2017deeplab} & TPAMI18 & .226 & .404 & .428 & .132 & .065 & .214 & .173 & .071 & .203 & .119 & .091 & .193\\
    \midrule
    MFCN \cite{salloum2018image} & JVCIP18 & .230 & .172 & .351 & .118 & .062 & .165 & .145 & .090 & .152 & .119 & {.102} & .155 \\ 
    \midrule 		
    RRU-Net \cite{bi2019rru} & CVPRW19 & .190 & .228 & .292 & .068 & .028 & .154 & .054 & .041 & .155 & .094 & .087 &  .126 \\ 
    \midrule						
    MantraNet \cite{wu2019mantra} & CVPR19 & .107 & .156 & .120 & .134 & .061 & .118 & .090 & .020 & .157 & .038 & .087 & .099\\ 	
    \midrule 
    HPFCN \cite{li2019localization} & ICCV19 & .155 & .074 & .180 & .069 & .028 & .094 & .052 & .047 & .093 & .081 & .068 & .086\\ 
    \midrule
    H-LSTM \cite{bappy2019hybrid} & TIP19 & \textbf{.354} & .140 & .140 & .130 & .044 & .187 & .114 & .053 & .155 & .131 & .133 & .144\\ 
    \midrule 								
    SPAN \cite{hu2020span} & ECCV20 & .195 & .454 & .153 & .142 & .031 & .141 & .077 & .046 & .166 & .075 & .023 & .137\\
    \midrule 	
    ViT-B \cite{dosovitskiy2020vit} & ICLR21 & .242 & .193 & .285 & .114 & .052 & .196 & .151 & .053 & .185 & \underline{.163} & .099 & .158\\	
    \midrule	
    Swin-ViT \cite{liu2021swin} & ICCV21 & .208 & .321 & .392 & .159 & .158 & {.303} & .175 & {.098} & .260 & .136 & .039 & .204\\	
    \midrule
    PSCC \cite{liu2022pscc} & TCSVT22 & .131 & .338 & .319 & .110 & .056 & .166 & {.184} & .035 & .156 & .085 & .046 & .148 \\ 
    \midrule 
    MVSS-Net++ \cite{dong2022mvss} & TPAMI22 & \underline{.289} & {.545} & {.503} & \textbf{.464} & {.097} & .265 & .170 & .068 & {.265} & .105 & .063 & {.258}\\	
    \midrule	
    CAT-NET \cite{kwon2022learning} & IJCV22 & .023 & .055 & .147 & .135 & \textbf{.216} & .208 & .125 & .043 & .109 & .040 & .042 & .104\\
    \midrule 											
    EVP \cite{liu2023evp} & CVPR23 & .205 & .266 & .478 & .103 & .090 & .236 & .055 & .082 & .228 & .118 & .096 & .178 \\	
    \midrule 
    \textcolor{black}{TruFor} \cite{guillaro2023trufor} & CVPR23 & .257 & \textbf{.795} & \underline{.536} & \underline{.270} & .147 & \underline{.358} & \underline{.210} & \underline{.117} & \underline{.344} & .117 & \underline{.149} & \underline{.300}  \\
    \midrule[1.5pt]
    \cellcolor[HTML]{E0DBDB}PIM & \cellcolor[HTML]{E0DBDB}Ours & \cellcolor[HTML]{E0DBDB}.264 & \cellcolor[HTML]{E0DBDB}\underline{.630} & \cellcolor[HTML]{E0DBDB}\textbf{.565} & \cellcolor[HTML]{E0DBDB}{.230} & \cellcolor[HTML]{E0DBDB}\underline{.162} & \cellcolor[HTML]{E0DBDB}\textbf{.415} & \cellcolor[HTML]{E0DBDB}\textbf{.229} & \cellcolor[HTML]{E0DBDB}\textbf{.142} & \cellcolor[HTML]{E0DBDB}\textbf{.396} & \cellcolor[HTML]{E0DBDB}\textbf{.228} & \cellcolor[HTML]{E0DBDB}\textbf{.212} & \cellcolor[HTML]{E0DBDB}\textbf{.318} \\
    \bottomrule
  \end{tabular}}
  \label{mcc_table}
\end{table*}

\newpage
\appendix
\vspace{-2mm}
\noindent \textcolor{black}{\textbf{Details of PIDA.} In our work, we exclusively use the CASIAv2 dataset for training, which includes 7,491 real images and 5,123 fake images. Only real images from CASIAv2 are used for Pixel-Inconsistency Data Augmentation (PIDA). Fig.~\ref{Selfblending} illustrates the PIDA pipeline. We apply four common perturbation types to the pristine real images ($I_{p}$): Gaussian blurriness, compression, noise, and color channel shuffling. By combining the corrupted image ($I_{c}$), the pristine real images ($I_{p}$), and the foreground mask ($M$), we generate the augmented forgery sample ($I_{b}$) and the corresponding label ($M$). For Gaussian blurriness, each $I_{p}$ in CASIAv2 is blurred with a kernel size $\in$ \{3, 5, 7, 9, 11\}. Each $I_{p}$ is also compressed with a random Quality Factor (QF) $\in$ [71, 95], and the standard deviation $\sigma$ of the Gaussian noise is randomly sampled from $\sigma$ $\in$ (0.01, 0.20). Additionally, we randomly shuffle the RGB color channels of $I_{p}$ to obtain $I_{c}$. Consequently, we obtain 7,491 $\times$ 4 PIDA forgery images. Each image is randomly horizontally flipped before being passed to the model during training. The purpose of PIDA is to drive the model to focus on extracting inherent pixel-level inconsistencies rather than semantic-level inconsistencies.}

\noindent \textbf{Additional evaluations.} In Table~\ref{auc_table}, we report the cross-dataset forgery localization performance using the threshold-free metric AUC. Notably, our method achieves an outstanding 78.2\% AUC performance. Compared with MVSS-Net++ \cite{dong2022mvss}, the proposed method achieves a 2.8$\%$ average AUC-score improvement, increasing  from 75.4$\%$ to 78.2$\%$. In Table~\ref{mcc_table}, our method consistently achieves the best or second-best detection performance on unseen testing datasets. And our average MCC performance outperforms SOTA methods by a clear margin. 

\begin{figure*}[ht]
\centering
\includegraphics[scale=0.32]{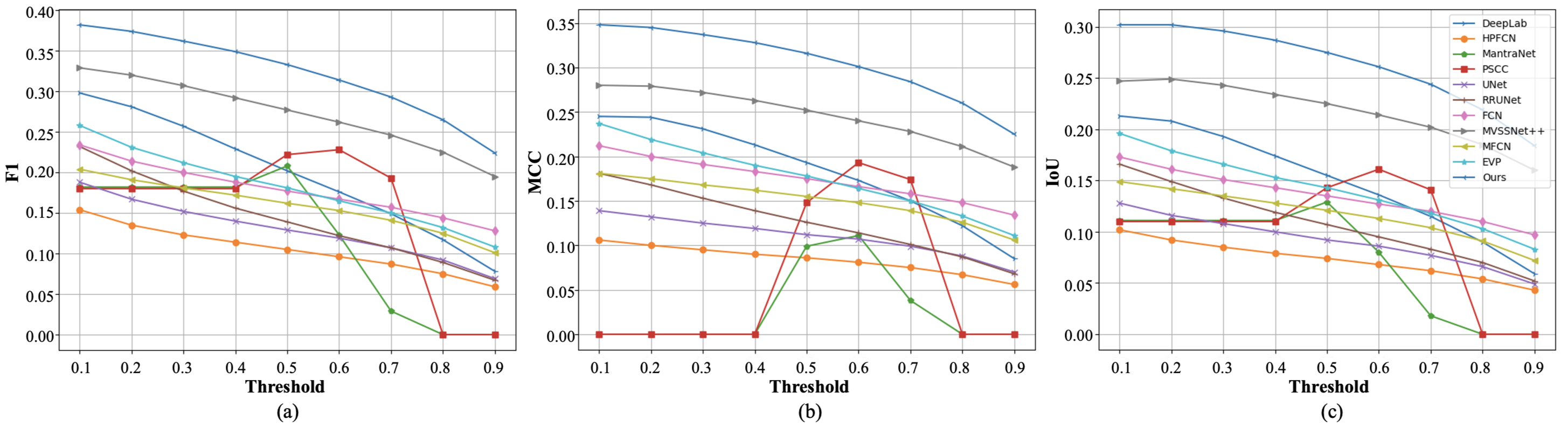}
\vspace{-4mm}
\caption{Image forgery localization performance with varying thresholds. (a). F1; (b). MCC; (c). IoU.}
\vspace{-4mm}
\label{threshold}
\end{figure*}

\noindent\textbf{Pixel-level evaluation at different thresholds.} The determination of threshold values is crucial for the final localization performance \cite{dong2022mvss}. We assess the effectiveness of our model's forgery localization across a range of threshold values from 0.1 to 0.9. We classify a pixel as a forgery if its predicted probability exceeds the specified threshold. Fig.~\ref{threshold} presents the average localization performance on the 11 unseen datasets using F1, MCC, and IoU metrics. Namely, we plot the average results under the cross-dataset setting with varying thresholds. Our proposed method consistently outperforms existing models across all thresholds, underscoring its superiority regardless of the threshold selection.

We observe that most detectors' performance continuously decreases with higher threshold values. This phenomenon may be attributed to subtle artifacts in challenging forgery regions, where detectors struggle to make confident decisions, resulting in reduced true positives (TP) at higher thresholds. This finding indicates the importance of selecting a lower threshold when deploying a forgery detector in real-world scenarios.

\noindent \textcolor{black}{\textbf{IoU score on the unseen manipulation: Inpainting.} We further report Image manipulation localization performance (IoU score with fixed threshold: 0.5) on the unseen Inpainting data in Table~\ref{iou_ip}. Our designed method PIM consistently achieves outstanding IoU scores across different inpainting techniques. Furthermore, PIM significantly overperforms the best model CAT-NET, demonstrating our method's superior generalizability to unseen manipulations from another perspective. }

\begin{figure*}[ht]
\centering
\includegraphics[scale=0.33]{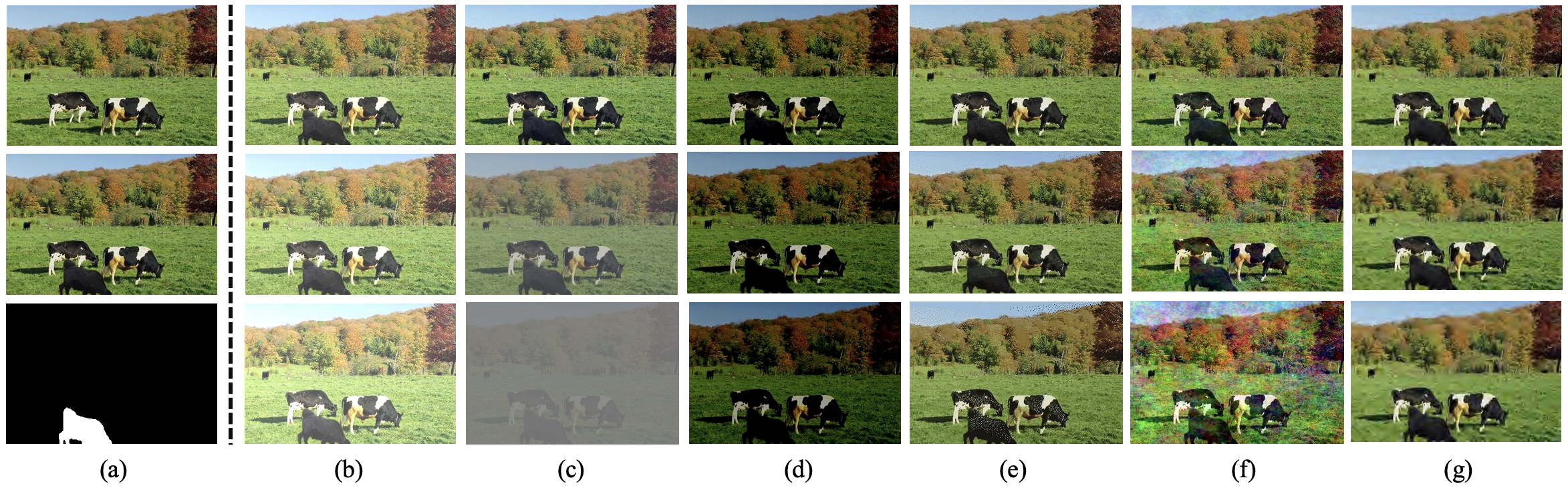}
\vspace{-4mm}
\caption{Showcases of (a). raw real image, raw forgery image, and ground-truth mask. The corresponding six image perturbation types of the raw forgery image: (b). Brightness; (c). Contrast; (d). Darkening; (e). Dithering; (f). Pink noise; (g). JPEG2000 compression. The top, middle, and bottom rows show Severity `1', `5', and `9' for all perturbation types.}
\vspace{-4mm}
\label{perturbation}
\end{figure*}

\noindent \textcolor{black}{\textbf{Showcases of perturbed images for robustness evaluation.} To mimic uncontrollable real-world scenarios, we incorporate six common image perturbation types with nine severity levels to examine the robustness of the image forgery localization models. The showcase examples of Severity `1', `5', and `9' are shown in Fig.~\ref{perturbation}.}

\begin{table}[]
  \caption{\textcolor{black}{Ablation experiments on MHSA.}}
  \vspace{-2mm}
  \label{MHSA}
\centering
\scalebox{0.78}{\begin{tabular}{cccc}
\toprule
 LPDE & GPDE & Avg. F1 & Avg. IoU \\ \hline
 MH & MH & \textbf{.333} & \textbf{.275} \\
\midrule[1.5pt]
 SH & MH & .318 & .270 \\
\hline
MH & SH & .312 & .261 \\
\hline
SH & SH & .298 & .245 \\
\bottomrule
\end{tabular}} 							
\end{table}

\begin{table}[]
  \caption{\textcolor{black}{Selection of loss weights.}}
  \vspace{-2mm}
  \label{Loss_weights}
\centering
\scalebox{0.8}{
\begin{tabular}{ccccc}
\toprule
$\lambda_B$ & $\lambda_C$ & $\lambda_R$ & F1 & AUC \\ \hline
1.0 & 0.01 & 0.1 & .320 & .752 \\ \hline	
1.0 & 0.1 & 0.1 & .312 & .740 \\ \hline
1.0 & 0.001 & 0.1 & \textbf{.333}  & \textbf{.782} \\ \hline
1.0 & 0.001 & 1.0 & .331 & .775 \\ \hline
1.0 & 0.001 & 0.01 & .327 & .769 \\ \bottomrule
\end{tabular}} 							
\end{table}

\begin{figure}[ht]
\centering
\includegraphics[scale=0.30]{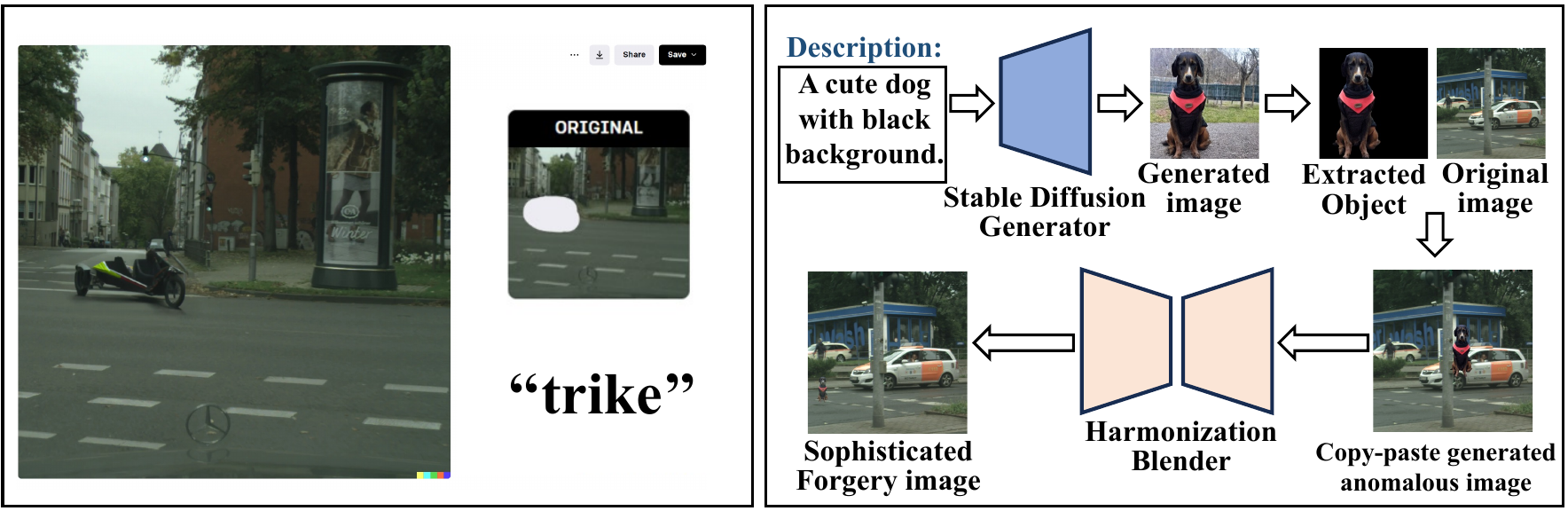}
\vspace{-6mm}
\caption{\textcolor{black}{Generation pipelines of Sophisticated manipulation pipelines. (a). Dall-E2 cityscape manipulation dataset. (b). Stable Diffusion cityscape manipulation dataset.}
}
\vspace{-4mm}
\label{sophiscticated}
\end{figure}

\noindent \textcolor{black}{\textbf{Generation details of the sophisticated datasets.} Fig.~\ref{sophiscticated} illustrates the generation pipelines of the two sophisticated datasets. In Fig.~\ref{sophiscticated} (a), we manually select the appropriate position and size for the generated object and then pass a reasonable object prompt to Dall-E2 (DE2) to obtain a photo-realistic image with a high level of harmonization. Since manually generating sophisticated forgery images is costly, we further apply existing algorithms in Fig.~\ref{sophiscticated} (b) to automatically produce sophisticated forgery images. The resulting images exhibit high-level harmonization, with the forgery object having compatible illumination, reasonable size, semantic consistency, and appropriate position. Consequently, the DE2 and SD datasets include 60 and 328 sophisticated fake images, respectively.}

\begin{figure*}[ht]
\centering
\includegraphics[scale=0.418]{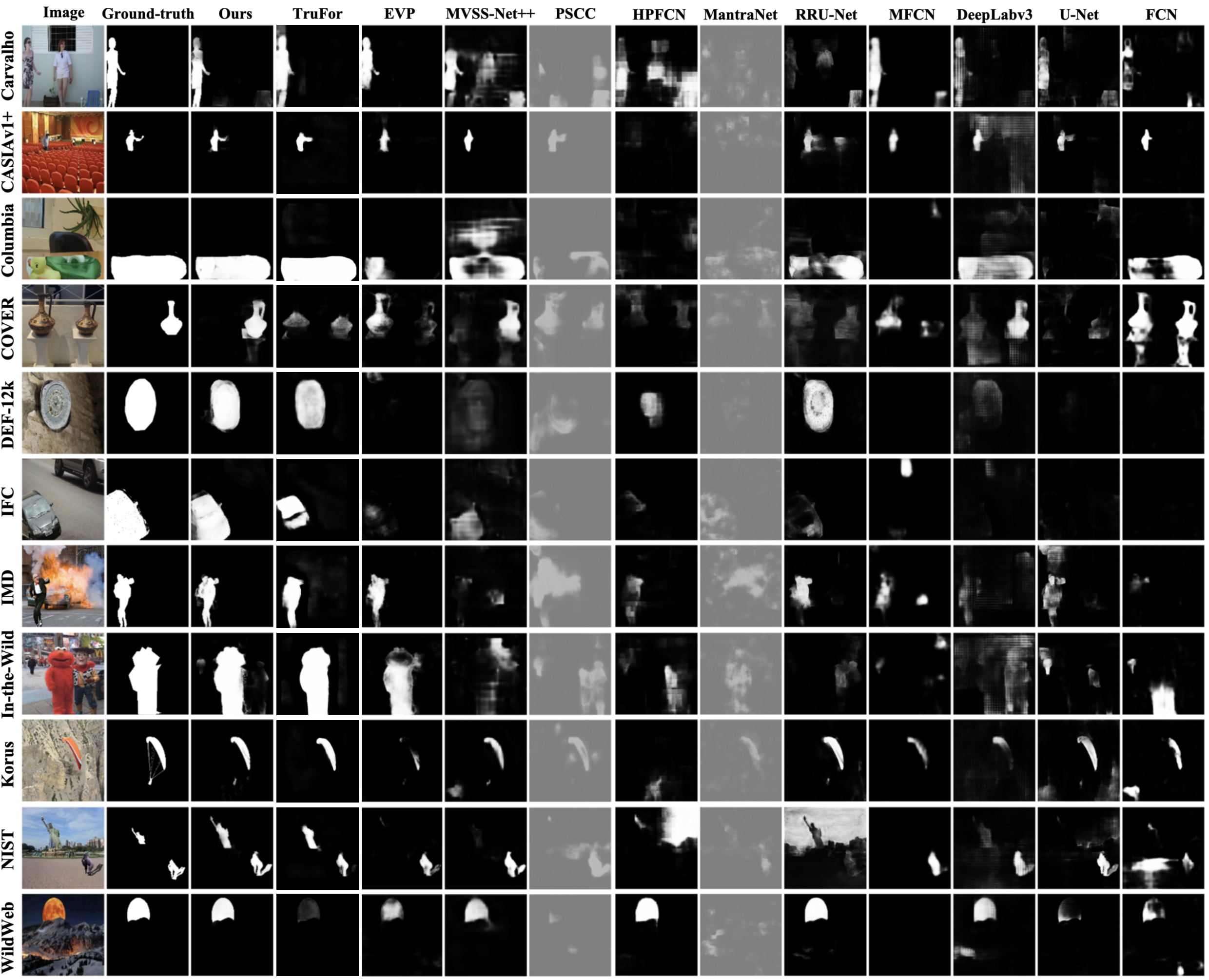}
\caption{\textcolor{black}{Additional forgery localization results on the 11 unseen test sets. The three left columns show the input images, corresponding ground-truth, and the localization results of our method. The right 11 columns present the results of SOTA methods.}
}
\label{visualization2}
\end{figure*}

\noindent \textbf{Additional visualization results.}
Fig.~\ref{visualization2} shows additional forgery localization results under the cross-dataset experimental setting. Our method accurately identifies the manipulated regions. In comparison with state-of-the-art (SOTA) methods, the proposed method demonstrates a superior forgery localization performance. 

\begin{table}[]
  \caption{\textcolor{black}{Trained on DEF-84k dataset.}}
  \vspace{-2mm}
  \label{transfer_def}
\centering
\scalebox{0.8}{
\begin{tabular}{ccccc}
\hline
\multirow{2}{*}{Method} &  \multicolumn{2}{c}{DEF-12k} & \multicolumn{2}{c}{CASIAv1+} \\ \cline{2-5}
& F1 & IoU & F1 & IoU \\ \hline	
Swin-ViT \cite{liu2021swin} & .477 & .423 & .058 & .048 \\ \hline
TruFor \cite{guillaro2023trufor} & .514 & .456 & .152 & .087 \\ \hline
\cellcolor[HTML]{E0DBDB}PIM (Ours) & \cellcolor[HTML]{E0DBDB}\textbf{.542} & \cellcolor[HTML]{E0DBDB}\textbf{.483} & \cellcolor[HTML]{E0DBDB}\textbf{.168} & \cellcolor[HTML]{E0DBDB}\textbf{.102} \\ \hline 
\end{tabular}} 							
\end{table}

\noindent \textcolor{black}{\textbf{Detailed ablation experimental results.} To enhance clarity, we present the detailed ablation experimental results (F1 and IoU scores) across the 11 testing datasets in Table~\ref{f1_table_abl} and Table~\ref{iou_table_abl}. The 12 listed settings indicate the use of different designed components, details of which can be found in Table~\ref{ablation}. The overall localization performances further demonstrate the effectiveness of the designed components.} \textcolor{black}{Compared to Setting 1 which only uses the transformer backbone (BB) to perform forgery localization, Setting 2 incorporates a boundary decoder, achieving a superior performance, particularly on the more challenging DEF-12k, IFC, and In-the-Wild datasets. This improvement highlights the significance of boundary information in enhancing the model's generalizability to practical scenarios. Setting 3 and 4 employ regular data augmentation (RDA) and the proposed pixel-inconsistency data augmentation (PIDA), respectively, further improving overall forgery localization performance. Notably, PIDA outperforms RDA in F1 and IoU scores across 8 out of 11 datasets, demonstrating its effectiveness. Furthermore, combining RDA and PIDA in Setting 5 yields additional performance gains, as the joint use of these augmentations enables the model to better handle complex forgeries. Setting 6 and 7 introduce central difference convolution (CDC) and radial difference convolution (RDC), respectively. Both modules consistently enhance performance across most datasets, as they effectively model local pixel dependencies critical for generalized forgery localization. In Setting 8, the naive concatenation of CDC and RDC features increases the diversity of captured local pixel-inconsistency features, resulting in overall improvements. To further optimize feature fusion, Setting 9 incorporates a learning-to-weight module (LWM), which dynamically adjusts the weights of CDC and RDC features based on different input images. This strategy significantly enhances generalizability on unseen datasets. Setting 10 integrates a compactness loss $L_{C}$, which delivers notable improvements on challenging CASIAv1+ and COVER datasets, likely due to their compact forgery regions. Setting 11 introduces a global pixel dependency encoder (GPDE), which significantly boosts F1 and IoU scores on the Columbia and Carvalho datasets with large forgery regions. This demonstrates that the proposed GPDE successfully models long-range pixel inconsistencies. However, relying heavily on GPDE causes slight performance drops on datasets with small manipulated regions. To address this limitation, Setting 12 employs a reconstruction loss, $L_{R}$, which encourages the model to also pay attention to image contents. This regularization effectively mitigates the issue, leading to overall performance enhancements.} 

\noindent \textcolor{black}{\textbf{Impacts of MHSA on image forgery localization.} 
In our designed model, we adopt the Multi-Head Self-Attention (MHSA) strategy in both the Local Pixel Dependency Encoder (LPDE) and the Global Pixel Dependency Encoder (GPDE), using head numbers of [3, 6, 12, 24] across the four transformer blocks. To examine the impact of MHSA, we conduct ablation experiments in Table.~\ref{MHSA}, where SH and MH refers to the Single-Head and Multi-Head Self-Attention mechanism, respectively. We report the average F1 and IoU scores across 11 unseen datasets in Table.~\ref{MHSA}. MHSA effectively scales the model's capacity and enables the model to search in larger feature space, resulting in superior image forgery localization performance compared to SHSA. In addition, it is observed that MHSA has a greater impact on GPDE than on LPDE. The potential reason could be that accurately modeling global pixel dependency for input images requires larger feature space. } 

\noindent \textcolor{black}{\textbf{Impacts of loss weights.} Table~\ref{Loss_weights} shows the average F1 and AUC across all 11 test datasets using different loss weights. We first fix $\lambda_B$ at 1.0, assigning equal importance to mask and boundary predictions. We then initialize $\lambda_C$ and $\lambda_R$ at 0.01 and 0.1, respectively, to balance the scale of the loss components in the early iterations. Subsequently, we tune $\lambda_C$ and $\lambda_R$ and report the image forgery localization results in Table~\ref{Loss_weights}. The model achieves the highest F1 and AUC scores on unseen datasets when $\lambda_C$ is 0.001 and $\lambda_R$ is 0.1.}

\textcolor{black}{Our trained model using the determined loss weights has been demonstrated effective on multiple forgery image datasets. The proposed method achieves strong generalizability across unseen traditional forgery datasets (IFC, CASIAv1+, WildWeb, COVER, NIST2016, Carvalho, Korus, In-the-wild, DEF-12k-test, and IMD2020), unseen inpainting datasets (CA, EC, GC, LB, LR, NS, PM, RN, SG, SH, and TE), and recent AIGC datasets (Dall-E2 (DE2), Stable Diffusion (SD), Autosplice, and CocoGlide). These experimental results verify the adaptability of the selected loss weights from another point of view.}

\textcolor{black}{To further validate our method's adaptability, we train our model on the DEF-84k image manipulation dataset \cite{mahfoudi2019defacto} using the same loss weights and compare it with previous methods, as shown in Table~\ref{transfer_def}. Note that all listed methods are trained on DEF-84k to ensure a fair comparison. It can be observed that our method PIM still achieves the best performance on the DEF-12k test set and best generalizability to the unseen CASIAv1+ dataset, demonstrating the adaptability of the selected loss weights.} 

\begin{figure*}[ht]
\centering
\includegraphics[scale=0.24]{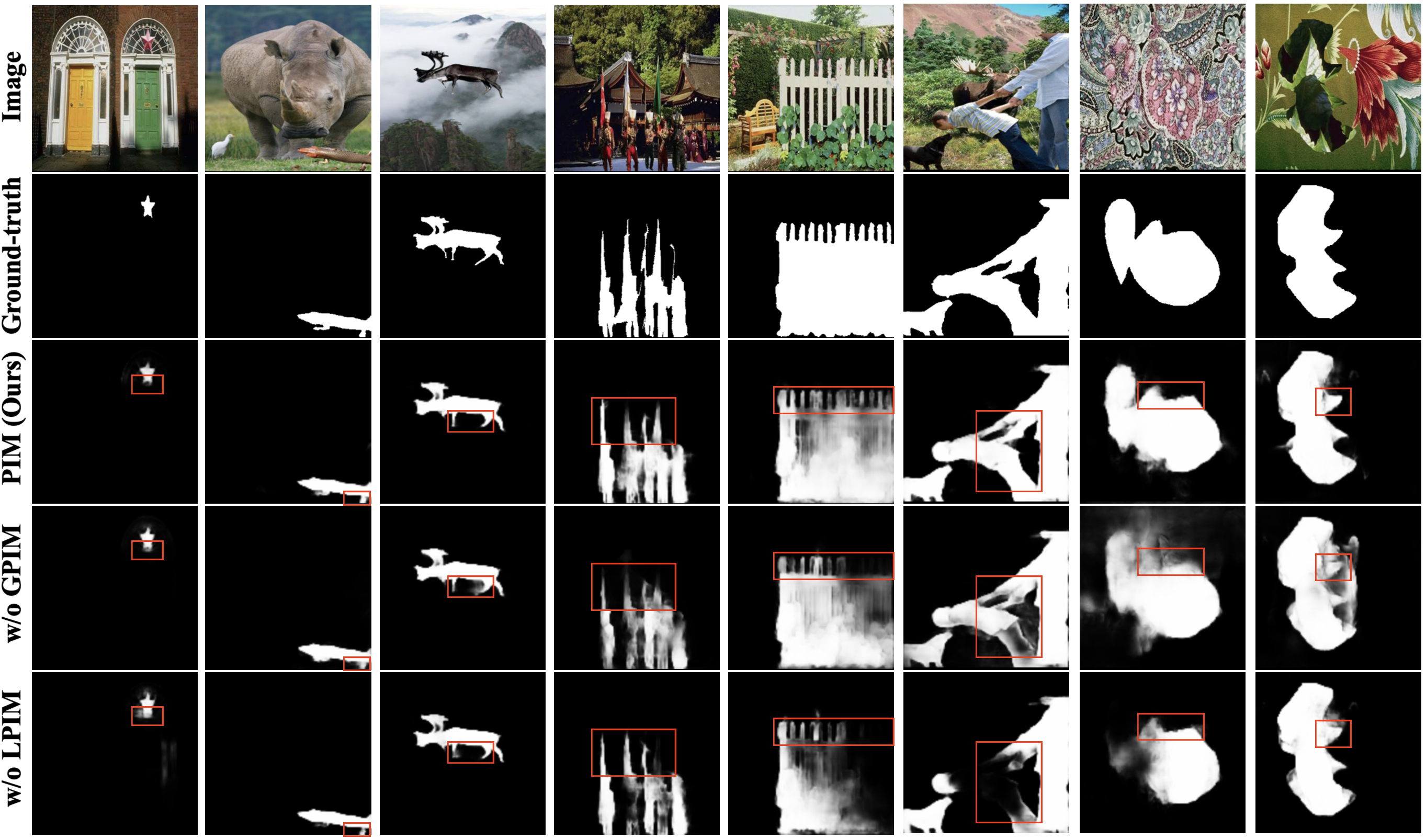}
\caption{\textcolor{black}{Visualization ablation experiments on the designed Global Pixel-Inconsistency Modeling (GPIM) and Local Pixel-Inconsistency Modeling (LPIM).}}
\label{PIM_ABL}
\end{figure*}

\noindent \textcolor{black}{\textbf{Visualization ablation experiments on GPIM \& LPIM.} 
To demonstrate the efficacy of the Global Pixel-Inconsistency Modeling (GPIM) and Local Pixel-Inconsistency Modeling (LPIM) strategies, we visualize the ablation results of image forgery localization maps in Fig.~\ref{PIM_ABL}. The top three rows represent the input images, the corresponding ground-truth masks, and the predicted results of our proposed Pixel-Inconsistency Modeling (PIM) method. The fourth row presents the predicted forgery maps without using the raster-scan mask in the attention mechanism, while the bottom row shows the results without the designed difference convolutions in the local pixel dependency encoder. From the highlighted red boxes, we observe that our proposed PIM method can more accurately localize forgery pixels, regardless of whether the forgery regions are substantial or subtle. This finding evidences that PIM indeed benefits from the designed GPIM and LPIM strategies, thereby achieving superior pixel-level forgery detection performance.}




\begin{table*}
  \centering
  \caption{\textcolor{black}{Image manipulation localization performance (\textbf{IoU score} with fixed threshold: 0.5) on the unseen manipulation type: Inpainting.}}
  \scalebox{0.80}{\begin{tabular}{lccccccccccccc}
    \toprule	
    Method & Venue & CA & EC & GC & LB & LR & NS & PM & RN & SG & SH & TE & AVG \\
    \midrule 
    FCN \cite{long2015fully} & CVPR15 & .065 & .024 & .005 & .019 & .385 & .083 & .161 & .095 & .243 & .083 & .037 & .109 \\
    \midrule 										
    U-Net \cite{ronneberger2015u} & MICCAI15 & .006 & .008 & .004 & .003 & .267 & .448 & .073 & .047 & .048 & .030 & .422 & .123 \\
    \midrule 											
    DeepLabv3 \cite{chen2017deeplab} & TPAMI18 & .075 & .050 & .006 & .015 & .467 & .541 & .180 & .144 & .378 & .095 & .493 & .222 \\
    \midrule  											
    MFCN \cite{salloum2018image} & JVCIP18 & .008 & .013 & .001 & .009 &  .136 & .490 & .031 & .048 & .032 & .046 & .480 & .118 \\
    \midrule  											
    RRU-Net \cite{bi2019rru} & CVPRW19 & .022 & .037 & .017 & .014 & .356 & .433 & .129 & .071 & .128 & .054 & .357 & .147 \\
    \midrule 											
    MantraNet \cite{wu2019mantra} & CVPR19 & .182 & .308 & \underline{.183} & .269 & .037 & .335 & .023 & .201 & .058 & .262 & .259 & .192\\ 
    \midrule 											
    HPFCN \cite{li2019localization} & ICCV19 & .006 & .007 & .004 & .005 & .120 & .400 & .013 & .026 & .012 & .021 & .365 & .089 \\
    \midrule  										
    H-LSTM \cite{bappy2019hybrid} & TIP19 & .028 & .018 & .024 & .021 & .076 & .034 & .025 & .038 & .022 & .028 & .028 & .031 \\
    \midrule  
    SPAN \cite{hu2020span} & ECCV20 & .005 & .022 & .006 & .004 & .289 & .361 & .077 & .085 & .146 & .012 & .184 & .108 \\
    \midrule 											
    ViT-B \cite{dosovitskiy2020vit} & ICLR21 & .012 & .010 & .009 & .019 & .075 & .281 & .012 & .023 & .020 & .033 & .270 & .069 \\
    \hline 											
    Swin-ViT \cite{liu2021swin} & ICCV21 & .158 & .177 & .003 & .057 & .308 & .144 & .304 & .245 & .277 & .221 & .039 & .176 \\
    \midrule  											
    PSCC \cite{liu2022pscc} & TCSVT22 & .208 & .207 & .066 & .126 & .186 & .519 & .112 & .181 & .226 & .148 & .479 & .223 \\
    \midrule 											
    MVSS-Net++ \cite{dong2022mvss} & TPAMI22 & .063 & .036 & .007 & .016 & \underline{.489} & \underline{.735} & .229 & .189 & .329 & .153 & \underline{.731} & .271 \\
    \midrule  											
    CAT-NET \cite{kwon2022learning} & IJCV22 & \underline{.450} & \underline{.429} & \textbf{.286} & \underline{.658} & .244 & .354 & .167 & \textbf{.426} & \underline{.470} & \underline{.509}  & .361 & \underline{.396} \\
    \midrule 
    EVP \cite{liu2023evp} & CVPR23 & .207 & .290 & .035 & .300 & .393 & .212 & \underline{.245} & .267 & .393 & .434 & .206 & .271 \\
    \midrule  											
    TruFor \cite{guillaro2023trufor} & CVPR23 & .119 & .102 & .105 & .210 & .119 & .126 & .042 & .102 & .064 & .079 & .125 & .108 \\
    \midrule
    \cellcolor[HTML]{E0DBDB}PIM & \cellcolor[HTML]{E0DBDB}Ours & \cellcolor[HTML]{E0DBDB}\textbf{.530} & \cellcolor[HTML]{E0DBDB}\textbf{.567} & \cellcolor[HTML]{E0DBDB}.052 & \cellcolor[HTML]{E0DBDB}\textbf{.702} & \cellcolor[HTML]{E0DBDB}\textbf{.690} & \cellcolor[HTML]{E0DBDB}\textbf{.758} & \cellcolor[HTML]{E0DBDB}\textbf{.416} & \cellcolor[HTML]{E0DBDB}\underline{.370} & \cellcolor[HTML]{E0DBDB}\textbf{.832} & \cellcolor[HTML]{E0DBDB}\textbf{.523} & \cellcolor[HTML]{E0DBDB}\textbf{.782} & \cellcolor[HTML]{E0DBDB}\textbf{.566} \\ 
    \bottomrule	 										
  \end{tabular}}
  \label{iou_ip}
\end{table*}

\begin{table*}
  \centering
  \caption{\textcolor{black}{Ablation study for image manipulation localization (\textbf{F1 score} with fixed threshold: 0.5).}}
  \scalebox{0.80}{\begin{tabular}{lccccccccccccc}
    \toprule 	
    Setting & NIST & Columbia & CASIAv1+ & COVER & DEF-12k & IMD & Carvalho & IFC & In-the-Wild & Korus & WildWeb & AVG \\
    \midrule 	
    1 & .220 & .365 & .390 & .168 & .157 & .300 & .183 & .102 & .265 & .134 & .040 & .211\\	
    \midrule 											
    2 & .186 & .339 & .358 & .124 & .166 & .346 & .251 & .112 & .336 & .168 & .036 & .220\\	
    \midrule 											
    3 & .233 & .485 & .515 & .132 & .126 & .281 & .175 & .132 & .267 & .166 & .046 & .233 \\	
    \midrule 											
    4 & .275 & .433 & .331 & .260 & .127 & .308 & .177 & .110 & .387 & .201 & .249 & .260 \\	
    \midrule 									
    5 & .237 & .614 & .525 & .175 & .168 & .380 & .129 & .137 & .372 & .200 & .177 & .283 \\	
    \midrule 											
    6 & .272 & .628 & .506 & .235 & .168 & .392 & .211 & .138 & .414 & .199 & .177 & .304\\	
    \midrule 											
    7 & .290 & .636 & .525 & .221 & .164 & .388 & .200 & .142 & .406 & .192 & .225 & .308 \\	
    \midrule 										
    8 & .294 & .670 & .526 & .222 & .169 & .393 & .174 & .135 & .418 & .213 & .220 & .312\\	
    \midrule  											
    9 & .269 & .754 & .507 & .235 & .162 & .395 & .232 & .149 & .408 & .183 & .194 & .317\\	
    \midrule 
    10 & .264 & .677 & .543 & .282 & .171 & .405 & .239 & .160 & .404 & .210 & .200 & .323\\
    \midrule 
    11 & .284 & .720 & .516 & .286 & .142 & .393 & .312 & .141 & .426 & .201 & .207 & .330\\
    \midrule
    12 & .280 & .680 & .566 & .251 & .167 & .419 & .253 & .155 & .418 & .234 & .236 & .333 \\
    \bottomrule
  \end{tabular}}
  \label{f1_table_abl}
\end{table*}

\begin{table*}
  \centering
  \caption{\textcolor{black}{Ablation study for image manipulation localization (\textbf{IoU score} with fixed threshold: 0.5).}}
  \scalebox{0.80}{\begin{tabular}{lccccccccccccc}
    \toprule 
    Setting & NIST & Columbia & CASIAv1+ & COVER & DEF-12k & IMD & Carvalho & IFC & In-the-Wild & Korus & WildWeb & AVG \\
    \midrule 
    1 & .167 & .297 & .356 & .124 & .129 & .243 & .132 & .078 & .214 & .103 & .033 & .171 \\	
    \midrule 										
    2 & .150 & .271 & .328 & .092 & .136 & .281 & .186 & .088 & .268 & .131 & .026 & .178 \\	
    \midrule 									
    3 & .180 & .395 & .471 & .098 & .101 & .226 & .135 & .104 & .212 & .128 & .035 & .190 \\
    \midrule 									
    4 & .222 & .367 & .281 & .200 & .101 & .242 & .126 & .085 & .305 & .161 & .208 & .209 \\	
    \midrule 								
    5 & .190 & .542 & .474 & .138 & .138 & .316 & .094 & .109 & .305 & .158 & .145 & .237\\	
    \midrule  									
    6 & .216 & .552 & .459 & .179 & .137 & .324 & .154 & .108 & .340 & .158 & .144 & .252\\
    \midrule 								
    7 & .235 & .562 & .475 & .179 & .133 & .319 & .147 & .112 & .332 & .154 & .188 & .258 \\
    \midrule 								
    8 & .242 & .598 & .473 & .178 & .136 & .324 & .129 & .109 & .339 & .172 & .183 & .262 \\	
    \midrule 							
    9 & .229 & .692 & .461 & .190 & .132 & .329 & .179 & .119 & .335 & .146 & .165 & .271 \\
    \midrule 										
    10 & .216 & .602 & .490 & .223 & .139 & .332 & .184  & .126 & .321 & .164 & .167 & .269\\
    \midrule 
    11 & .228 & .642 & .465 & .226 & .110 & .319 & .229 & .106 & .343 & .155 & .167 & .272\\
    \midrule 
    12 & .225 & .604 & .512 & .188 & .133 & .340 & .194 & .119 & .338 & .182 & .193 & .275 \\
    \bottomrule 
  \end{tabular}}
  \label{iou_table_abl}
\end{table*}

\end{document}